\documentclass[aps,prd,reprint,superscriptaddress,nofootinbib,
               preprintnumbers,floatfix]{revtex4-2}
\pdfoutput=1

\makeatletter
\let\bibliography\bibliography@latex
\auto@bib@empty
\makeatother

\usepackage{amsmath}
\usepackage{amssymb}
\usepackage{array}
\usepackage{booktabs}
\usepackage{ragged2e}
\usepackage{booktabs}
\usepackage{siunitx}
\usepackage{calc}
\usepackage[colorlinks=true,
            linkcolor=blue,
            citecolor=blue,
            urlcolor=blue]{hyperref}
\usepackage{orcidlink}
\usepackage{multirow}
\usepackage{nicefrac}
\usepackage{cleveref}
\usepackage{pstricks}
\usepackage{graphicx}
\graphicspath{{figures/}}
\usepackage{xspace}
\usepackage{listings}
\usepackage{algorithm}
\usepackage{algpseudocode}
\usepackage{
  pgf,
  tikz}
\usetikzlibrary{
  patterns,
  shapes.multipart,
  arrows,
  trees,
  scopes,
  decorations.pathreplacing,
  decorations.pathmorphing,
  decorations.markings,
  decorations.text,
  positioning,
  calc
}
\usepackage[compat=1.1.0]{tikz-feynman}

\usepackage{subfig}
\usepackage{placeins}
\usepackage{relsize}
\usepackage{todonotes}
\setuptodonotes{fancyline, inline, inlinewidth=\linewidth, color=blue!30}
\usepackage{newunicodechar}
\newunicodechar{−}{-}

\newcommand{\myfigure}[5][h]{
  \begin{figure}[#1]
    \begin{center}
      \includegraphics[width=#2\textwidth]{#3}\\[0.5em]
      \parbox[t]{#2\textwidth}{
        \caption{#4}
        \ifx&#5&\else\label{#5}\fi
      }
    \end{center}
  \end{figure}
}

\newcommand{\done}{{\mathrm d}}
\newcommand{\ddone}{{\mathrm d}}

\newcommand{\oforder}[1]{\ensuremath{\mathcal{O}\left(#1 \right)}\xspace}

\newcommand{\bea}{\begin{eqnarray}}
\newcommand{\eea}{\end{eqnarray}}
\newcommand{\beq}{\begin{equation}}
\newcommand{\eeq}{\end{equation}}

\newcommand{\bs}{\begin{split}}
\newcommand{\es}{\end{split}}

\newcommand{\bi}{\begin{itemize}}
\newcommand{\ei}{\end{itemize}}
\newcommand{\bc}{\begin{center}}
\newcommand{\ec}{\end{center}}
\newcommand{\bac}{\begin{array}{c}}
\newcommand{\bacc}{\begin{array}{cc}}
\newcommand{\ea}{\end{array}}

\def\spa#1.#2{\langle#1\,#2\rangle}
\def\spb#1.#2{[#1\,#2]}

\renewcommand\Re{\operatorname{\mathcal{R}e}}

\newcommand{\UMeV}{\ensuremath{\mathrm{MeV}}\xspace}
\newcommand{\UGeV}{\ensuremath{\mathrm{GeV}}\xspace}

\newcommand{\mrad}{\ensuremath{\mathrm{mrad}}\xspace}

\newcommand{\Virtual}{\ensuremath{{\mathcal{V}}}}

\newcommand{\NLOEW}{\ensuremath{\mathrm{NLO_{EW}}}\xspace}
\newcommand{\NNLOEW}{\ensuremath{\mathrm{NNLO_{EW}}}\xspace}
\newcommand{\NNLO}{\ensuremath{\mathrm{NNLO}}\xspace}

\newcommand{\sla}[1]{\ensuremath{{#1\kern-0.45em/}}}

\newcommand{\eik}[1]{\ensuremath{\tilde{S}\left(k_{#1}\right)}}

\newcommand{\EqRef}[1]{Eq.~\eqref{#1}\xspace}

\newcommand{\EqRefs}[2]{Eqs.~\eqref{#1} and \eqref{#2}\xspace}

\newcommand{\SecRef}[1]{Sec.~\ref{#1}\xspace}
\newcommand{\SecRefC}[1]{Section~\ref{#1}\xspace}

\newcommand{\AppRef}[1]{Appendix~\ref{#1}\xspace}

\newcommand{\FigRef}[1]{Fig.~\ref{#1}\xspace}
\newcommand{\FigRefs}[2]{Figs.~\ref{#1} and~\ref{#2}\xspace}

\newcommand{\TabRef}[1]{Table~\ref{#1}\xspace}

\newcommand\lep{L\scalebox{0.8}{EP}\xspace}

\newcommand\LEP{\lep}

\newcommand{\kkmc}{KKMC\xspace}

\newcommand{\OpenLoops}{O\protect\scalebox{0.8}{PEN}L\protect\scalebox{0.8}{OOPS}\xspace}
\newcommand{\Recola}{R\protect\scalebox{0.8}{ECOLA}\xspace}
\newcommand{\Griffin}{G\protect\scalebox{0.8}{RIFFIN}\xspace}
\newcommand{\BabaYaga}{B\protect\scalebox{0.8}{ABA}Y\protect\scalebox{0.8}{AGA}\xspace}
\newcommand{\McMule}{M\protect\scalebox{0.8}{C}M\protect\scalebox{0.8}{ULE}\xspace}

\newcommand{\Sherpa}{S\protect\scalebox{0.8}{HERPA}\xspace}
\newcommand{\Comix}{C\protect\scalebox{0.8}{OMIX}\xspace}

\newcommand{\Rivet}{R\protect\scalebox{0.8}{IVET}\xspace}
\newcommand{\YODA}{Y\protect\scalebox{0.8}{ODA}\xspace}

\DeclareOldFontCommand{\rm}{\normalfont\rmfamily}{\mathrm}
\DeclareOldFontCommand{\sf}{\normalfont\sffamily}{\mathsf}
\DeclareOldFontCommand{\tt}{\normalfont\ttfamily}{\mathtt}
\DeclareOldFontCommand{\bf}{\normalfont\bfseries}{\mathbf}
\DeclareOldFontCommand{\it}{\normalfont\itshape}{\mathit}
\DeclareOldFontCommand{\sl}{\normalfont\slshape}{\@nomath\sl}
\DeclareOldFontCommand{\sc}{\normalfont\scshape}{\@nomath\sc}

\usepackage[T1]{fontenc}

\newcommand{\CEEX}{CEEX\xspace}
\newcommand{\EEX}{EEX\xspace}
\newcommand{\OCEEX}[1]{\ensuremath{\mathcal O(\alpha^{#1})_{\rm CEEX}}\xspace}
\newcommand{\YFS}{YFS\xspace}
\newcommand{\ACRAIC}{A\protect\scalebox{0.8}{CRAIC}\xspace}
\newcommand{\YFSNLOEW}{\ensuremath{\text{YFS}_{\mathbf{NLO_{EW}}}}\xspace}
\newcommand{\YFSLO}{\ensuremath{\text{YFS}_{\mathbf{LO}}}\xspace}
\newcommand{\YFSnNLO}{\ensuremath{\text{YFS}_{\mathbf{nNLO_{EW}}}}\xspace}
\newcommand{\hsum}{\mathop{\overline{\sum}}}
\newcommand{\avsq}[1]{\ensuremath{\overline{|#1|^2}}}
\newcommand{\BG}{Berends--Giele\xspace}

\newcommand{\Mamp}{\ensuremath{\mathfrak{M}}}
\newcommand{\Mzero}{\ensuremath{\Mamp_0}}
\newcommand{\Mn}[1]{\ensuremath{\Mamp_{#1}}}
\newcommand{\Coh}{\ensuremath{\mathcal{C}}}
\newcommand{\sfac}{\ensuremath{\mathfrak{s}}}
\newcommand{\Sins}{\ensuremath{\mathbb{S}}}
\newcommand{\bet}[1]{\ensuremath{\hat\beta_{#1}}}
\newcommand{\betsq}[1]{\ensuremath{\tilde\beta_{#1}}}
\newcommand{\Yfac}{\ensuremath{Y}}
\newcommand{\Om}{\ensuremath{\Omega}}
\newcommand{\hel}{\ensuremath{\{\lambda\}}}

\newcommand{\stage}{\ensuremath{g}}
\newcommand{\Nstage}{\ensuremath{G}}
\newcommand{\prt}{\ensuremath{\wp}}
\newcommand{\Gres}{\ensuremath{\mathcal G}}

\newcommand{\newinline}[1]{{\color{blue}#1}}

\definecolor{stageI}{RGB}{0,90,170}
\definecolor{stageF}{RGB}{190,40,30}
\colorlet{photonI}{black}
\colorlet{photonF}{black}
\newcommand{\stagecolours}{\colorlet{photonI}{stageI}\colorlet{photonF}{stageF}}
\newcommand{\sI}{\textcolor{stageI}{\sfac_{\mathrm{I}}}}
\newcommand{\sF}{\textcolor{stageF}{\sfac_{\mathrm{F}}}}

\newcommand{\gdia}[2]{
\begin{tikzpicture}[baseline={([yshift=-0.6ex]current bounding box.center)},scale=0.55,transform shape]
\begin{feynman}
\vertex (i1) at (0,1.15) {\Large$e^-$};
\vertex (i2) at (0,-1.15) {\Large$e^+$};
\vertex [draw, circle, thick, fill=gray!18, inner sep=0pt, minimum size=1.55cm] (h) at (1.8,0) {\Large$#2$};
\vertex (o1) at (3.5,1.25) {\Large$1$};
\vertex (o2) at (3.6,0.45) {\Large$2$};
\node at (3.35,-0.28) {\Large$\vdots$};
\vertex (oN) at (3.5,-1.15) {\Large$N$};
\ifcase#1
  \diagram*{ (i1) -- (h) -- (i2),
             (h) -- (o1), (h) -- (o2), (h) -- (oN) };
\or
  \vertex (m) at (0.6,0.76); \vertex (g) at (1.3,1.75); \node[photonI] at (1.5,1.95) {\Large$k$};
  \diagram*{ (i1) -- (m) -- (h) -- (i2),
             (m) -- [photon, photonI, thick] (g),
             (h) -- (o1), (h) -- (o2), (h) -- (oN) };
\or
  \vertex (m) at (0.6,-0.76); \vertex (g) at (1.3,-1.75); \node[photonI] at (1.5,-1.95) {\Large$k$};
  \diagram*{ (i1) -- (h) -- (m) -- (i2),
             (m) -- [photon, photonI, thick] (g),
             (h) -- (o1), (h) -- (o2), (h) -- (oN) };
\or
  \vertex (m) at (2.8,0.74); \vertex (g) at (2.35,1.85); \node[photonF] at (2.2,2.1) {\Large$k$};
  \diagram*{ (i1) -- (h) -- (i2),
             (h) -- (m) -- (o1), (m) -- [photon, photonF, thick] (g),
             (h) -- (o2), (h) -- (oN) };
\or
  \vertex (m) at (2.8,-0.68); \vertex (g) at (2.35,-1.8); \node[photonF] at (2.2,-2.05) {\Large$k$};
  \diagram*{ (i1) -- (h) -- (i2),
             (h) -- (o1), (h) -- (o2),
             (h) -- (m) -- (oN), (m) -- [photon, photonF, thick] (g) };
\fi
\end{feynman}
\end{tikzpicture}}
\newcommand{\legsum}[2]{\gdia{1}{#1} + \gdia{2}{#1} + \gdia{3}{#2} + \cdots + \gdia{4}{#2}}

\begin{document}

\title{Automation of amplitude-level QED resummation with \texorpdfstring{\ACRAIC}{ACRAIC}}
\author{Alan~Price\,\orcidlink{0000-0002-0372-1060}}
\email{alan.price@uj.edu.pl}
\affiliation{Jagiellonian University, ul.\ prof.\ Stanis\l{}awa \L{}ojasiewicza 11,
30-348 Krak\'{o}w, Poland}

\preprint{MCNET-26-33}

\begin{abstract}
  Precision electroweak predictions for future $e^+e^-$ colliders will require
levels of precision unmatched in collider physics and the modelling of
multiple-photon emissions, and the exponentiation of the associated logarithms
needs to be carefully considered. The Yennie--Frautschi--Suura theorem subtracts
the infrared singularities to all orders and exponentiates them, leaving
infrared-finite remainders which can be calculated order-by-order. Coherent
 exponentiation performs this subtraction at the amplitude level,
preserving the interference between emissions, but has so far been available
only for fermion-pair production, with hand-coded amplitudes. We present
\ACRAIC, an automated, amplitude-level realization of this subtraction for
general scattering processes in \Sherpa. This method allows us to
construct an \oforder{\alpha} and \oforder{\alpha^2} amplitude level
subtraction scheme, in which the infrared-finite remainders are tree-level
helicity amplitudes with partition-dependent propagator shifts, computed by the
 \BG recursion, and the one-loop electroweak virtual enters through a
per-helicity factor from the loop provider. At \oforder{\alpha^2} two
independent matchings agree on the size of the missing higher orders to
$\sim10^{-3}$, and the estimated perturbative uncertainty shrinks by about an
order of magnitude relative to \oforder{\alpha}.
\end{abstract}

\maketitle
\section{Introduction}\label{sec:intro}
The physics programs of the current and proposed lepton
colliders~\cite{Anashkin:2006zz, Adinolfi:2002uk, BaBar:2001yhh, Abbiendi:2016xup, BESIII:2009fln, Achasov:1999ju, ILC:2013jhg,Aicheler:2012bya,FCC:2018evy,CEPCStudyGroup:2018ghi,Vernieri:2022fae}
rest on a theoretical control of QED radiation that is at least as good as the
experimental systematics~\cite{Jadach:2019bye,Blondel:2018mad}. QED corrections
to the initial and final states distort every observable that is sensitive to
photon emissions, at the level of tens of percent, so that a description
accurate to the per-mil level is a precondition for interpreting the data
rather than a refinement of it.

The organizing principle for all-order QED is the theorem of Yennie, Frautschi
and Suura (\YFS)~\cite{Yennie:1961ad}. The infrared (IR) singular parts of the
real and virtual corrections are extracted to all orders and exponentiated,
leaving IR-finite remainders that are computed order by order and retain the
full kinematic dependence of each emission. Two equivalent realizations of the
theorem are in use. The first is the so-called exclusive exponentiation (\EEX)
where the factorization is performed on the squared amplitude and the remainders
are squared quantities and the sum over which charged line emitted a given
photon is incoherent, with coherent effects added as a corrective weight. While
in the coherent exclusive exponentiation
(\CEEX)~\cite{Jadach:1999vf,Jadach:2000ir} the factorization is performed on
the amplitude, the remainders are complex helicity amplitudes, and the sum over
the assignments of photons to emitters is taken before squaring, so that the
interference between initial-state radiation (ISR) and final-state radiation
(FSR), the initial-final interference (IFI), is kept at the amplitude level.
\CEEX was derived and realized in \kkmc~\cite{Jadach:1999vf,Jadach:2022mbe} for
$e^+e^-\to f\bar f$, and was carried over to $q\bar q\to f\bar f$ in hadron
collisions in \kkmc-hh~\cite{Jadach:2016zsp}. The same family of \YFS Monte
Carlos was instrumental throughout the \LEP physics program, for the
luminosity measurement~\cite{Jadach:1996is}, the $Z$-peak
lineshape~\cite{Jadach:1988gb}, lepton-pair~\cite{Jadach:1999tr} and
$W$-pair~\cite{Jadach:1997llk} production. Other codes such as
\BabaYaga~\cite{CarloniCalame:2000pz} and \McMule~\cite{Aliberti:2024fpq} reach
comparable precision for specific final states, the former through a
structure-function resummation matched to a QED parton shower and the latter
at fixed \NNLO order.

In general-purpose event
generators~\cite{Alwall:2014hca,Kilian:2007gr,Sherpa:2024mfk,Bellm:2025pcw,Bierlich:2022pfr,Frixione:2021zdp,Campbell:2022qmc},
multi-photon radiation beyond fixed order is supplied by a universal QED
parton
shower~\cite{Sjostrand:1985xi,Corcella:2000bw,Barberio:1990ms,CarloniCalame:2007cd,Flower:2026byh},
accurate to the leading logarithm. The fixed-order electroweak corrections
themselves are now automated for generic
processes~\cite{Schonherr:2017qcj,Frederix:2018nkq}, but matching them to a
shower in the POWHEG/MC@NLO sense~\cite{Frixione:2007vw,Frixione:2002ik} has
not been realized for $e^+e^-$ colliders, nor would it retain the
amplitude-level coherence of \CEEX beyond the first emission. A third,
collinear-factorization approach convolutes an electron parton distribution
function~\cite{Skrzypek:1990qs}, known to next-to-leading logarithmic
accuracy~\cite{Bertone:2019hks} and recently to two
loops~\cite{Stahlhofen:2025hqd,Schnubel:2025ejl}, with a short-distance
partonic cross section. This inclusive picture does not by itself give the
fully exclusive, coherent photon kinematics that \CEEX and \ACRAIC provide.

\Sherpa~\cite{Bothmann:2019yzt,Sherpa:2024mfk} computes tree-level helicity
amplitudes for essentially arbitrary Standard Model final states with
\Comix~\cite{Gleisberg:2008fv}, a \BG recursion~\cite{Berends:1987me}, obtains
one-loop amplitudes from \OpenLoops~\cite{Buccioni:2019sur} and
\Recola~\cite{Actis:2016mpe,Denner:2017wsf} through its
interfaces~\cite{Biedermann:2017yoi}, and carries \YFS machinery for
decays~\cite{Schonherr:2008av} and for lepton
colliders~\cite{Krauss:2022ajk,Price:2025fiu}. The latter provides a
squared-level \YFS exponentiation matched to the complete one-loop electroweak
correction, denoted \YFSNLOEW as in~\cite{Price:2025fiu}. What has been missing
is the amplitude-level form of the resummation at arbitrary multiplicity. The
need shows at higher multiplicity, where the squared-level matching can become
unstable while the amplitude-level alternative remains stable, as we find for
four-lepton production in \SecRef{sec:4f}.

In this paper we present \ACRAIC\footnote{ACRAIC is an acronym for ``A Coherent
    Resummation of Amplitudes In C++''.}, an automated amplitude-level \YFS
exponentiation for general scattering processes, built on the \Comix recursion
inside \Sherpa. Its construction rests on three observations. First, the
ingredients of \CEEX other than the coherent sum, the \YFS form factor, the
gauge invariance of the total soft factor and the IR finiteness, depend on the
process only through the charges and momenta of the external legs. Second, the
coherent sum generalizes from the initial/final assignment to an assignment of
photons to \emph{emission stages}, and requiring each stage to be separately
gauge invariant fixes an admissible decomposition to be a partition of the
charged legs into charge-neutral sets, which we derive from the radiating
dipoles of the process, including resonance stages for $Z/\gamma^*$ and $W$
pairs. Third, the remainders are tree-level helicity amplitudes with $N+2$ and
$N+3$ legs evaluated with partition-dependent propagator momenta, which a \BG
recursion computes directly, provided it meets a small set of requirements on
gauge conventions and numerical accuracy that squared-level applications never
expose. The one-loop virtual enters through a per-helicity factor built from
the loop provider's amplitudes. We validate \ACRAIC against \kkmc for
$e^+e^-\to\mu^+\mu^-$ in cross sections and distributions at photonic $\mathcal
    O(\alpha^2)$, the second order without the two-loop $\gamma Z$ boxes and weak
corrections, and compare it with \YFSNLOEW, which shares the resummation and
differs only in the matching, for multiple processes at various energies. This
paper is mostly technical, and we illustrate the resulting machinery with two
example applications, the Higgs recoil mass at FCC-ee and the $Z$-pole
precision distributions, which a dedicated phenomenological study could develop
further, and with a stability test at higher multiplicity in four-lepton
production (\SecRef{sec:4f}).

The paper is organized as follows. \SecRefC{sec:yfs} recalls the \YFS form
factor and the squared and amplitude-level soft factors.
\SecRefC{sec:algorithm} describes the \ACRAIC algorithm, namely the stages,
partitions and propagator shifts, the remainders, the crude density and the
weight, the treatment of soft photons, the virtual correction and the cost.
\SecRefC{sec:realisation} states what the construction requires of an automated
amplitude generator and how \Comix meets it. \SecRefC{sec:validation} presents
the validation, and \SecRef{sec:conclusions} concludes. The conventions for the
soft factors are collected in \AppRef{app:conventions}.

\section{\texorpdfstring{\YFS}{YFS} exponentiation}\label{sec:yfs}
This section fixes the notation and recalls the ingredients of \YFS
exponentiation that are used below. Standard results are quoted without
derivation and for explicit details we refer to~\cite{Yennie:1961ad,Jadach:2000ir,Krauss:2022ajk}. We follow the notation of~\cite{Price:2025fiu} and every expression is written for an arbitrary
number of external legs.

\label{sec:kinematics}

We consider a process with two incoming momenta $p_a,p_b$ and $N$ outgoing
momenta, and treat all $N+2$ external legs as one list $\{p_i\}$, each leg
carrying a charge $Q_i$ in units of the positron charge and a direction label
\begin{equation}
  \theta_i =
  \begin{cases}
    -1 & \text{incoming}, \\
    +1 & \text{outgoing}.
  \end{cases}
  \label{eq:theta}
\end{equation}
Charge conservation is the single statement
\begin{equation}
  \sum_{i} Q_i\,\theta_i = 0 .
  \label{eq:charge-conservation}
\end{equation}
The photons are $k_1,\dots,k_n$ with helicities $\sigma_j=\pm1$.
The IR-singular parts of the virtual corrections and of the real emissions
inside the soft-photon domain $\Om$, bounded by the IR cutoff,
exponentiate. As in Ref.~\cite{Price:2025fiu} they combine
into the \YFS form factor
\begin{equation}
  \Yfac(\Om) = \sum_{i<j}
  \Big[ \Re \mathcal B_{ij} + \tilde{\mathcal B}_{ij}(\Om) \Big],
  \label{eq:yfs-formfactor}
\end{equation}
in which the virtual function $\mathcal B_{ij}$ and the real function
$\tilde{\mathcal B}_{ij}(\Om)$, each including its factor $2\alpha$, depend only
on the momenta and charges of the dipole $ij$. The IR poles cancel dipole by
dipole, and the number of external legs enters only as the range of the sum.
\Sherpa builds \EqRef{eq:yfs-formfactor} from its initial-initial,
initial-final and final-final dipoles~\cite{Krauss:2022ajk,Schonherr:2008av},
so the form factor is available at arbitrary multiplicity.
The squared-level soft factor of a real photon, the eikonal factor, is written
as a single current with its normalization explicit,
\begin{equation}
  \tilde S(k) = -\frac{\alpha}{4\pi^2}\,
  \Big(\sum_i Q_i\theta_i\,\frac{p_i}{p_i\cdot k}\Big)^{2},
  \label{eq:stilde}
\end{equation}
with the photon phasespace weight $\done^3k/k^0$. The cross section is then
\begin{align}
  \ddone\sigma^{\left(\infty\right)} & = \sum_{n_\gamma=0}^\infty
  \frac{e^{Y(\Omega)}}{n_\gamma!}\,
  \ddone{\Phi_n}
  \left[\prod_{i=1}^{n_\gamma}\ddone{\Phi_i^\gamma}\,\eik{i}\,\Theta(k_i,\Omega)\right] \nonumber \\
                                     & \quad\times\Bigg(\tilde{\beta}_0\left(\Phi_n\right)
  + \sum_{j=1}^{n_\gamma}\frac{\tilde{\beta}_{1}(\Phi_{n+1})}{\eik{j}} \nonumber                  \\
                                     & \qquad\quad+\sum_{{j,k=1}\atop{j< k}}^{n_\gamma}
  \frac{\tilde{\beta}_{2}(\Phi_{n+2})}{\eik{j}\eik{k}}
  + \cdots\Bigg)\,,\label{eq:eex}
\end{align}
where $ \ddone{\Phi_n}$ is the phase space of the hard final state and the
virtual orders are suppressed,
\begin{equation}
  \betsq{n_R}=\sum_{n_V}\betsq{n_R}^{n_V+n_R} ,
  \label{eq:beta-order-sum}
\end{equation}
with the superscript the order in $\alpha$ for the given correction.

\label{sec:softfactor}

For a single real photon of momentum $k$ and helicity $\sigma$ the eikonal
current at the level of the amplitude is given by
\begin{equation}
  \sfac^{(\sigma)}(k)
  = e \sum_{i} Q_i\,\theta_i\,
  \frac{\varepsilon^{(\sigma)*}\!\cdot p_i}{k\cdot p_i} .
  \label{eq:soft-factor}
\end{equation}
It is a sum over single legs, and the pairwise structure of
\EqRefs{eq:yfs-formfactor}{eq:stilde} arises only after squaring,
\begin{equation}
  \tilde S(k) = \sum_\sigma|\sfac^{(\sigma)}(k)|^2/(16\pi^3) ,
  \label{eq:stilde-amplitude}
\end{equation}
matching \EqRef{eq:stilde}.

At the amplitude level, we can write the exact Feynman amplitude with $n$
photons, $\Mn{n}$, as a coherent sum over the assignments $\prt$ of the photons
to emitters. For $e^+e^-\to f\bar f$ each photon is assigned to the initial (I)
or the final (F) state,
\begin{equation}
  \begin{split}
    \Mn{n}\big(\hel;\{k_j,\sigma_j\}\big)
    ={} & \sum_{\prt} S_\prt\,\bet{0}^\prt(\hel)
    \\  &
       + \sum_{j=1}^{n}\sum_{\prt_{\bar\jmath}} S_{\prt_{\bar\jmath}}\,
       \bet{1}^{\prt_{\bar\jmath}}(\hel;k_j,\sigma_j)
    \\ & + \dots ,
  \end{split}
  \label{eq:partition-sum}
\end{equation}
where $\hel$ collects the helicities of the external particles, $\prt$ runs
over $\{\mathrm I,\mathrm F\}^n$ and $\prt_{\bar\jmath}$ over the assignments
of the photons other than $j$, $S_\prt=\prod_i\sfac^{(\sigma_i)}_{\prt(i)}(k_i)$
is the product over the photons that $\prt$ assigns, $\sfac_{\mathrm I}$ and
$\sfac_{\mathrm F}$ are the parts of \EqRef{eq:soft-factor} from the initial
and final legs, and the
superscript $\prt$ marks an amplitude evaluated with its $s$-channel
propagator at the momentum the partition leaves for the hard process,
\begin{equation}
  P-K^\prt_{\mathrm I},\qquad
  K^\prt_{\mathrm I}=\sum_{j:\,\prt(j)=\mathrm I}k_j ,
  \label{eq:partition-momentum}
\end{equation}
with $P$ the momentum of the beams, $K^\prt_{\mathrm I}$ the photons the
partition assigns to the initial state, and $s_\prt=(P-K^\prt_{\mathrm I})^2$
the reduced invariant of the partition. In~\cite{Jadach:2000ir} the authors split
the remainder of photon $j$ into an initial- and a final-state part, one for
each assignment of that photon and \EqRef{eq:partition-sum} carries their
sum, which depends only on the assignment of the other photons and is what a
generated one-photon amplitude provides. The remainders are helicity
amplitudes, with the order in $\alpha$ in parentheses, $\bet{n}^{(r)}$, to
distinguish them from the squared-level $\betsq{n}^{r}$ and from the partition
superscript. At \NLOEW we have,
\begin{align}
  \bet{0}^{(0)}(\hel)          & = \Mzero(\hel), \label{eq:beta0}                                        \\
  \bet{0}^{(1)}(\hel)          & = \Mzero^{(1)}(\hel)
  - \tfrac12\,\mathcal B_{\mathrm V}\,\Mzero(\hel), \label{eq:beta0-1loop}                               \\
  \bet{1}^{(1)}(\hel;k,\sigma) & = \Mn{1}(\hel;k,\sigma)
  - \sfac^{(\sigma)}_{\mathrm I}(k)\,\Mzero^{\mathrm I}(\hel) \nonumber                                  \\
                               & \quad - \sfac^{(\sigma)}_{\mathrm F}(k)\,\Mzero(\hel), \label{eq:beta1}
\end{align}
where $\Mzero^{\mathrm I}$ is the Born with its $s$-channel propagator at
$P-k$, the photon being assigned to the initial state, $\Mzero^{(1)}$ is the
one-loop amplitude of the loop provider, and
$\mathcal B_{\mathrm V}$ is the virtual form factor at the level of the
amplitude. Its real part is the subtraction $\sum_{i<j}\Re\mathcal B_{ij}$ of
$\betsq{0}^{1}$. We write $\hsum_{\hel}$ for the
sum over helicities averaged over the initial-state ones, $\tfrac14\sum_{\hel}$
for unpolarized beams and with beam polarizations $P_{1,2}$ the initial-state
helicities are weighted by $(1\pm P_{1,2})/2$ instead. Squaring,
\begin{equation}
  \hsum_{\hel}\bigl|\bet{0}^{(0)}+\bet{0}^{(1)}\bigr|^2
  = \betsq{0}^{0}+\betsq{0}^{1}+\hsum_{\hel}\bigl|\bet{0}^{(1)}\bigr|^2.
  \label{eq:virtual-identity}
\end{equation}
The two matchings carry the same $\mathcal O(\alpha)$ virtual and differ only at
$\mathcal O(\alpha^2)$.
The cross section follows from \EqRef{eq:partition-sum} as in \EqRef{eq:eex},
with the helicity-averaged $\hsum_{\hel,\sigma}|\Mn{n}|^2$ in place of the
product of $\tilde S$ and the squared remainders.

The difference between the two matchings is where the modulus is taken, as
shown in \FigRefs{fig:yfsnloew}{fig:acraic}. \YFSNLOEW squares the complete
one-photon matrix element and subtracts the eikonal times the Born, photon by
photon. \ACRAIC forms $\bet{1}$ per helicity at the amplitude level, with the
resonance propagators moved according to the stage the photon is assigned to,
sums the partitions of all photons coherently and squares last.
\begin{figure*}
  \centering
  \input{figures/diagrams/fig-yfsnloew}
  \caption{The squared-level matching of the $\mathcal O(\alpha)$ real
    remainder for a generic process $e^+e^-\to N$. The shaded blob is the hard
    amplitude, legs $1\ldots N$ its final state, and a photon attaches to
    $e^\pm$ and to every charged final leg. The remainder
    $\betsq{1}^{1}$ is formed after squaring, the full one-photon matrix
    element squared minus $\tilde S(k)\,|\Mzero|^2$.}
  \label{fig:yfsnloew}
\end{figure*}

\begin{figure*}
  \centering
  \input{figures/diagrams/fig-acraic}
  \caption{The amplitude-level matching of the $\mathcal O(\alpha)$ real
    remainder for a given process.
    $\bet{1}(k)$ is formed per helicity before squaring. The subtraction is the soft limit of the same attachments, the
    stage soft factors times the Born at the stage's momentum,
    $\sfac_{\mathrm I}(k)$ at $P-k$ and $\sfac_{\mathrm F}(k)$ at $P$, with
    $\sfac_{\mathrm I}+\sfac_{\mathrm F}=\sfac$.
    The partitions $\prt$ of the photons into
    stages are summed coherently with the stage soft factors
    $\sfac_{\prt(j)}$, and only then is the result squared. The braces group
    the attachments of each stage with its own subtraction, so each is IR
    finite by itself.}
  \label{fig:acraic}
\end{figure*}

\section{The \texorpdfstring{\ACRAIC}{ACRAIC} algorithm}\label{sec:algorithm}
This section describes the \ACRAIC algorithm and how the coherent sum of
\EqRef{eq:partition-sum} is generalized to arbitrary final states, which
amplitudes it requires, and how the event weight is formed. What the algorithm
demands of the amplitude generator that supplies those amplitudes is the
subject of \SecRef{sec:realisation}.

\subsection{Emission stages and partitions}\label{sec:stages}

For a general $2\to N$ process the relevant question about a photon is not
whether it is initial- or final-state, but which propagator's virtuality it
changes. We therefore replace the initial/final assignment by an assignment to
\emph{emission stages}, sets $L_\stage$, $\stage=1,\dots,\Nstage$, of charged
legs, separated from one another by resonant internal propagators that a photon
may straddle. The stage-resolved soft factor is the part of
\EqRef{eq:soft-factor} from the legs of the stage,
\begin{equation}
  \sfac_{\stage}^{(\sigma)}(k) = e \sum_{i \in L_\stage}
  Q_i\,\theta_i\, \frac{\varepsilon^{(\sigma)*}\!\cdot p_i}{k\cdot p_i}.
  \label{eq:stage-soft-factor}
\end{equation}

The coherent sum is assembled from the individual $\sfac_\stage$, not from
their total, and under $\varepsilon^{(\sigma)}\to\varepsilon^{(\sigma)}+\lambda
  k$ each changes by
\begin{equation}
  \delta\,\sfac_{\stage}^{(\sigma)}(k)
  = e\,\lambda \sum_{i \in L_\stage} Q_i\,\theta_i ,
  \label{eq:stage-gauge-variation}
\end{equation}
a partial charge sum. Gauge invariance of every term of the coherent sum
therefore requires every stage to be charge neutral,
\begin{equation}
  \sum_{i \in L_\stage} Q_i\,\theta_i = 0
  \quad \text{for every stage } \stage ,
  \label{eq:neutrality}
\end{equation}
and the requirement that no emitter be omitted or counted twice is the closure
condition
\begin{equation}
  \sum_{\stage=1}^{\Nstage} \sfac_{\stage}^{(\sigma)}(k) = \sfac^{(\sigma)}(k).
  \label{eq:closure}
\end{equation}
The two conditions are independent and a charged pair assigned to two stages at
once leaves every stage neutral and violates \EqRef{eq:closure}. A
subprocess bounded by resonant propagators satisfies \EqRef{eq:neutrality}
provided the resonance is counted as a leg of both the stage that produces it
($\theta=+1$) and the stage in which it decays ($\theta=-1$).

\ACRAIC derives the stages from the dipoles that the \YFS generator of \Sherpa
radiates from. The initial state forms one stage and each
final-state dipole forms one more. A candidate stage that fails
\EqRef{eq:neutrality} is merged into a single final stage, which is then
neutral by \EqRef{eq:charge-conservation}. For $2\to2$ this gives $\Nstage=2$,
and \EqRef{eq:partition-sum} is recovered term by term with $\sfac_{\mathrm
    I}+\sfac_{\mathrm F}=\sfac$. A final state without charged particles has
$\Nstage=1$ and every photon belongs to the initial stage and the coherent sum has
a single term.

For a $Z/\gamma^*$ decaying to a charged pair the stage is the pair itself,
which is neutral without the resonance. For a charged resonance the resonance
must be a leg. When the generator identifies a $W^+W^-$ pair, \ACRAIC uses a
production stage $\{e^-,e^+,W^-,W^+\}$ and one decay stage per $W$, in which
the $W$ is an incoming leg with its momentum reconstructed from its decay
products and the $W$ terms cancel between production and decay in
\EqRef{eq:closure}. The stage structure is that of the double-pole treatment of
$W$-pair production, in which radiation is attributed to production and to each
decay~\cite{Jadach:1997llk}. It follows from the identity
\begin{equation}
  \frac{1}{D(P'+k)\,D(P')} = \frac{1}{2P'\!\cdot k}
  \left[\frac{1}{D(P')}-\frac{1}{D(P'+k)}\right],
  \label{eq:partial-fraction}
\end{equation}
with $D(p)=p^2-M_W^2+iM_W\Gamma_W$, for a photon emitted from the $W$ line of
momentum $P'$. The first term is the
photon on the production stage, with the $W$ propagator at $P'$, the second the
photon on the decay stage, with the $W$ propagator at $P'+k$, and each comes with
the corresponding eikonal factor of the $W$.
More than one decomposition can satisfy \EqRefs{eq:neutrality}{eq:closure} for
the same process. The $W$-pair final state admits the one-stage and the
three-stage decomposition. The admissible decompositions agree at $\mathcal
  O(\alpha)$ and differ beyond it and they are different resummation schemes, and
the difference between them is a theoretical uncertainty rather than a defect
of either. This follows from the construction itself, since both
decompositions satisfy \EqRefs{eq:neutrality}{eq:closure} and reproduce the
same $\mathcal O(\alpha)$ amplitude, not from a numerical comparison on the
$W$-pair process, where it is left for future work. A more sophisticated treatment of the radiation from the
quasi-stable $W$, along the lines of the QED exponentiation for quasi-stable
charged particles of~\cite{Jadach:2019wol}, is another possibility and we leave it to
future work.

\label{sec:partitions}

A partition $\prt$ assigns each photon to a stage and with $\Nstage$ stages
there are $\Nstage^n$ of them, and \EqRef{eq:partition-sum} becomes
\begin{equation}
  \begin{split}
    \Mn{n} ={} & \sum_{\prt} S_\prt\,\bet{0}^\prt
    + \sum_{j}\sum_{\prt_{\bar\jmath}} S_{\prt_{\bar\jmath}}\,
    \bet{1}^{\prt_{\bar\jmath}}(k_j)
    \\       &
            + \sum_{j<l}\sum_{\prt_{\bar\jmath\bar l}} S_{\prt_{\bar\jmath\bar l}}\,
            \bet{2}^{\prt_{\bar\jmath\bar l}}(k_j,k_l) + \dots ,
  \end{split}
  \label{eq:stage-partition-sum}
\end{equation}
where $\prt$ maps the photons onto $\{1,\dots,\Nstage\}$,
$\prt_{\bar\jmath}$ and $\prt_{\bar\jmath\bar l}$ map the photons other than
$j$, and other than $j$ and $l$, the soft factors $S$ are the products of the
stage soft factors of the photons each map assigns, and the superscript now
stands for the shifts of all internal momenta of the hard amplitude in that
partition. Let $K_\stage=\sum_{j:\,\prt(j)=\stage}k_j$
be the photon momentum the partition assigns to stage $\stage$, which
generalizes $K^\prt_{\mathrm I}$ of \EqRef{eq:partition-momentum}. The partition
Born $\bet{0}^\prt$ is evaluated with the external spinors and polarization
vectors at the physical momenta, and with the propagator momentum of every
internal line shifted as if the photons of each stage had been emitted from that
stage's legs and the $s$-channel line carries $P-K_{\mathrm
  I}$, with $K_{\mathrm I}$ the photons of the initial stage, and the resonance of
a final stage with legs of total momentum $q_\stage$ carries
$q_\stage+K_\stage$. In a recursion over off-shell currents,
the rule is formulated per current $c$ with leg set
$L(c)$ and every stage contained in $L(c)$ contributes its $K_\stage$ to the shift
of $c$, and a current whose leg set is larger than its complement also receives
the $K_\stage$ of the stages it contains in part, which keeps the pole of the
emitting line adjacent to the photon in place. For processes with a
spacelike exchange, the
propagator is set instead to the partition's reduced momentum transfer,
computed from the legs reduced by the partition's photons, which is exact in
both collinear limits.

The external momenta of the partition Born do not conserve momentum once the
photons are removed, and this is intended. The eikonal theorem expresses the
soft limit of $\Mn{1}$ through the Born at the physical external momenta, with
the internal lines adjacent to the emission carrying the photon. The stage Born
with shifted propagators is therefore the exact soft limit of the corresponding
assignment, as \EqRef{eq:partial-fraction} shows for a resonance.

\subsection{Remainders and virtual corrections}\label{sec:remainders}

We write $\Mzero(\hel;\{\Delta\})$ and $\Mn{1}(\hel;k,\sigma;\{\Delta\})$ for
the Born and one-photon tree amplitudes of \EqRefs{eq:beta0}{eq:beta1}, with
$N+2$ and $N+3$ legs, evaluated at the physical legs with a set $\{\Delta\}$ of
propagator shifts. A superscript $\prt$ stands for the shifts $\{\Delta_\prt\}$
of a partition. Thus $\bet{0}^\prt=\Mzero^\prt$ is the partition Born, and
$\Mn{1}^\prt(k_j)$ is the one-photon amplitude with $k_j$ an external leg and
the photons that $\prt$ assigns entering as shifts. For an assignment $\prt$
that leaves photon $i$ unassigned, $\prt_{i\to\stage}$ is $\prt$ extended by
photon $i$ on stage $\stage$, and the eikonal insertion of photon $i$,
\begin{equation}
  \Sins_i\,X^\prt = \sum_\stage \sfac_\stage(k_i)\,X^{\prt_{i\to\stage}} ,
  \label{eq:Sins}
\end{equation}
where $X^\prt$ stands for any amplitude or remainder carrying the partition
superscript of the previous paragraph, such as $\Mzero^\prt$ or
$\Mn{1}^\prt(k_j)$. The insertion multiplies an amplitude by the photon's
soft factor on each stage and places the photon there. Insertions of
different photons commute, and starting from no
assignment they build the partition sum, so that for a set $P$ of photons,
\begin{equation}
  \prod_{i\in P}\Sins_i\,X=\sum_{\prt_P}S_{\prt_P}X^{\prt_P}
\end{equation}
with $\prt_P$ the maps of $P$ onto the stages. From here on the superscript is left
implicit, and every amplitude and remainder is evaluated with the shifts of the
assignment that the insertions acting on it build up.
The one-photon remainder of \EqRef{eq:stage-partition-sum} is then
\begin{equation}
  \bet{1}(k_j) = \Mn{1}(k_j) - \Sins_j\,\bet{0} ,
  \label{eq:beta1-stage}
\end{equation}
the generalization of \EqRef{eq:beta1} in which each stage's Born sits at that
assignment's propagator kinematics.
The coherent amplitudes that \ACRAIC evaluates are
\EqRef{eq:stage-partition-sum} truncated after $\bet{0}$ and after $\bet{1}$.
We write them as $\Coh$ to
distinguish them from the Feynman amplitudes $\Mn{n}$. A $\Coh$ is built from
soft factors and the residuals $\bet{k}$ up to a fixed $k$, so it has the
soft limit of $\Mn{n}$ but differs from it by the residuals it omits, and it equals $\Mn{n}$ only when all residuals up to
$\bet{n}$ are kept. The $\mathcal O(\alpha^0)$ and $\mathcal O(\alpha^1)$ coherent amplitudes are
\begin{align}
  \Coh_0 ={} & \sum_\prt S_\prt\,\bet{0}^\prt ,
  \label{eq:A0}                                    \\
  \Coh_1 ={} & \Bigl[\prod_{i}\Sins_i\Bigr]\bet{0}
  + \sum_{j\in \mathcal{R}}\Bigl[\prod_{i\ne j}\Sins_i\Bigr]\bet{1}(k_j) ,
  \label{eq:A1}
\end{align}
where $\prt$ runs over the maps of the photons onto the stages, the
products over the photons, and  $\mathcal{R}$ is the set of photons whose remainder is
evaluated. $\mathcal{R}$ is taken by default to be all photons in an event, but it can be truncated to a subset of them, such
as the hardest $J$ photons, to reduce the number of hard amplitudes that the generator must evaluate.

Two properties of \EqRef{eq:A1} are used below. First, with one photon it
returns $\Coh_1=\Mn{1}(k)$ identically, for any Born and shift rule, since the Born term is $\Sins\,\bet{0}$ and the remainder subtracts exactly
that. The $\mathcal O(\alpha)$ real emission is therefore exact. Second, at
$n\ge2$ every partition Born $\bet{0}^\prt$ is added once and subtracted once
for every photon in $\mathcal{R}$, and the resulting $(1-|\mathcal{R}|)$ cancellation holds only if
the partition Born and the Born in the subtraction are one object, evaluated at
the same legs with the same shifts, and \EqRef{eq:A1} is written so that they
are.

The tree-level $\mathcal O(\alpha^2)$ remainder follows from the same
subtraction. For photons $j$ and $l$,
\begin{equation}
  \bet{2}(k_j,k_l) = \Mn{2}(k_j,k_l)
  - \Sins_l\,\bet{1}(k_j) - \Sins_j\,\bet{1}(k_l)
  - \Sins_j\Sins_l\,\bet{0} ,
  \label{eq:beta2}
\end{equation}
with one $N+4$ leg amplitude per pair of hard photons and assignment of the
others. The same subtraction defines the tree-level remainder of any number of
photons.
Truncating the remainders at $J$ photons means that every set $K$ of at most
$J$ photons enters as hard legs of a remainder, while all other photons are
inserted through their soft factors,
\begin{align}
  \Coh_J ={} & \sum_{K,\,|K|\le J}\Bigl[\prod_{i\notin K}\Sins_i\Bigr]\bet{|K|}(K)
  \nonumber                                                                        \\
  ={}        & \Bigl[\prod_{i}\Sins_i\Bigr]\bet{0}
  + \sum_{j}\Bigl[\prod_{i\ne j}\Sins_i\Bigr]\bet{1}(k_j)
  \nonumber                                                                        \\
             & + \sum_{j<l}\Bigl[\prod_{i\ne j,l}\Sins_i\Bigr]\bet{2}(k_j,k_l)
  + \dots ,
  \label{eq:AJ}
\end{align}
where the sum stops at remainders of $J$ photons. The first two terms are
\EqRef{eq:A1}, so $\Coh_1$ is the case $J=1$.
The truncation at $\bet{|K|}$ is suppressed in the soft limit of every photon
of $K$, so the terms left out at $J$ are suppressed by the energy fractions of
at least $J+1$ photons. In the implementation the sets are visited in order of
size within each partition and $J$ is a run-time choice.

\label{sec:virtual}

The IR-finite one-loop remainder enters as a per-helicity factor on the
Born-level amplitude $\Coh_0$ of \EqRef{eq:A0}, added to the $\Coh_1$ of
\EqRef{eq:A1} helicity by helicity,
\begin{equation}
  \Coh_1^V(\hel)=\Coh_1(\hel)+\Virtual(\hel)\,\Coh_0(\hel) .
  \label{eq:virt}
\end{equation}
With the tree and one-loop helicity amplitudes $a_0(\hel)$ and $a_1(\hel)$ of
the loop provider at the event's Born point,
\begin{equation}
  \Virtual(\hel)=\frac{a_1(\hel)}{a_0(\hel)}
  -\tfrac12\,\mathcal B_{\mathrm V},
  \label{eq:delta-hel}
\end{equation}
where $\mathcal B_{\mathrm V}$ is the virtual form factor of
\EqRef{eq:beta0-1loop}, evaluated in the coupling of the loop provider, so that $\Virtual(\hel)\,\Mzero(\hel)=\bet{0}^{(1)}(\hel)$. The
ratio $a_1/a_0$ cancels the provider's normalization and phase conventions, so
that only the map between its helicity labels and those of the tree generator
is needed. With $\bet{0}^{(1)}=\Virtual\,\Mzero$, the
$\mathcal O(\alpha)$ term of \EqRef{eq:virtual-identity} is the interference of
the one-loop remainder with the Born,
\begin{equation}
  \begin{split}
    \betsq{0}^{1}
     & =\hsum_{\hel}2\,\Re\bigl[\Mzero^*(\hel)\,\Virtual(\hel)\,\Mzero(\hel)\bigr]
    \\
     & =2\,\hsum_{\hel}|\Mzero(\hel)|^2\,\Re\,\Virtual(\hel) ,
  \end{split}
  \label{eq:delta-interference}
\end{equation}
in which $\Mzero^*\Mzero=|\Mzero|^2$ is real for each helicity, so only
$\Re\,\Virtual$ enters, whatever the phase of $\Virtual(\hel)$. Dividing by
$\betsq{0}^{0}=\hsum_{\hel}|\Mzero(\hel)|^2$, the Born-weighted average
\begin{equation}
  \bar{\Virtual}=\frac{\sum_{\hel}|\Mzero(\hel)|^2\,\Re\,\Virtual(\hel)}
  {\sum_{\hel}|\Mzero(\hel)|^2}
  =\frac{\betsq{0}^{1}}{2\,\betsq{0}^{0}}
  \label{eq:delta-bar}
\end{equation}
is half the squared-level virtual relative to the Born, so that on photon-free
events \ACRAIC and \YFSNLOEW carry the same $\mathcal O(\alpha)$ virtual.

The factor $\Virtual(\hel)$ is computed once per event, at the Born point, and
applied to every partition. A provider that returns only the helicity-summed
virtual supplies $\Virtual(\hel)=\bar{\Virtual}$ for all helicities, and
\EqRef{eq:virt} becomes
\begin{align}
  \avsq{\Coh_1^V}             & =\avsq{\Coh_1}
  +2\bar{\Virtual}\,\Re\langle\Coh_0,\Coh_1\rangle
  +\bar{\Virtual}^{\,2}\,\avsq{\Coh_0} ,
  \label{eq:ext-virtual}                       \\
  \langle\Coh_0,\Coh_1\rangle & =\hsum_{\hel}
  \Coh_0^\ast(\hel)\Coh_1(\hel) ,
  \label{eq:inner-product}
\end{align}
This helicity-averaged form is correct at $\mathcal O(\alpha)$ for unpolarized
cross sections and it averages the helicity dependence of the boxes. The
results of this paper apply the virtual instead to the whole $\mathcal
  O(\alpha)$ amplitude,
\begin{equation}
  \avsq{\Coh_1^V}=(1+\bar{\Virtual})^2\avsq{\Coh_1} ,
  \label{eq:ext-virtual-full}
\end{equation}
and the difference is of $\mathcal O(\alpha^2)$, the factorizable real-virtual,
but it can be restored at runtime.

Near a resonance the initial-final part of $\Virtual$ contains a large
logarithm of the resonance's off-shellness. For a final stage $\stage$ whose
legs a propagator of mass $M$ and width $\Gamma$ spans, it is
\begin{equation}
  \begin{aligned}
    I_\stage(q^2) & =a_\stage\ln\frac{\tilde M^2-q^2}{\tilde M^2},                 \\
    a_\stage      & =\frac{\alpha}{\pi}\sum_{i\notin L_\stage}\sum_{j\in L_\stage}
    Q_i\theta_i\,Q_j\theta_j\,\ln\bigl(2|p_i\cdot p_j|\bigr),
  \end{aligned}
  \label{eq:resonant-log}
\end{equation}
with $\tilde M^2=M^2-iM\Gamma$ and $q^2$ the invariant of the resonance. The
sum runs over the pairs of charged legs on the two sides of the stage's
boundary, and the scale of the logarithm cancels by \EqRef{eq:neutrality}. For
$e^+e^-\to f\bar f$,
\beq
a_\stage=-2(\alpha/\pi)Q_eQ_f\ln(t/u)
\eeq
and $I_\stage$ is the resonant part of the $\gamma Z$ box that \kkmc exponentiates into the $Z$
propagator following~\cite{Greco:1975rm}. A real photon with energy
$E\gg\Gamma$ does not interfere between the production and the decay of the
resonance, because its two partitions place the resonance at invariants
that are well separated. The coherent sum applies this suppression to every
amplitude, $\Coh_1$ included, whereas \EqRef{eq:virt} applies the
compensating virtual logarithm to $\Coh_0$ only. \ACRAIC
therefore removes $I_\stage$ from $\Virtual$ and multiplies the propagator of the
stage's resonance by $\exp[I_\stage(q^2_\prt)]$ in every amplitude of the
coherent sum, $\bet{0}$, each photon's $\Mn{1}$ and $\bet{2}$, each at the
invariant $q^2_\prt$ its partition gives that propagator. The subtraction is
exact helicity by helicity. Since the tree amplitude is linear in each
resonant propagator, evaluating it once with that propagator multiplied by
$1+I_\stage$ gives
\begin{equation}
  \Virtual_I(\hel)\,\Mzero(\hel)=I_\stage\,\Mzero^{\mathrm{res}}(\hel) ,
  \label{eq:resonant-delta}
\end{equation}
and $\Virtual\to\Virtual-\Virtual_I$. Everything is built from the stages and the
propagators the tree generator already has, so no process-specific box formula
enters.
A pair between two decay stages, as in $e^+e^-\to ZZ\to4\ell$, straddles two
resonances. Each stage then takes the pair's logarithm with its own resonance's
off-shellness, which reproduces each single-resonance logarithm but not the
mixed functions of the exact non-factorizable correction. For charged
resonances we do not exponentiate the logarithm of the decay stage, but leave
it to future work to implement the exponentiation of the non-factorizable
corrections of~\cite{Jadach:2019wol}.

The structure of \EqRef{eq:resonant-log} is established at one loop. In the
pole approximation the resonant soft-photon, non-factorizable, correction is
universal for any number of charged or neutral resonances. It depends only on
the charges, the momenta and the complex masses, with the photon momentum
routed through the resonance propagators as in the stages
here~\cite{Dittmaier:2015bfe}. The factorizable and non-factorizable parts
separate to all orders in the double-pole
approximation~\cite{Chapovsky:2001zt}, but an exponentiated form has been
derived only for a single resonance, at two loops~\cite{Ahmadova:2026ggg}, and
proposed for $W$ pairs~\cite{Jadach:2019wol}. Beyond $\mathcal O(\alpha)$ the
exponentiation of $I_\stage$ for several resonances is therefore an
extrapolation, exact at $\mathcal O(\alpha)$ and valid to leading power in
$\Gamma/M$.

\label{sec:oalpha2}

At $\mathcal O(\alpha^2)$ the real-virtuals need to be added, as the one-loop
correction to the emission of a photon, and the double real, the tree-level
emission of two.
\ACRAIC adds the tree-level $\bet{2}$ of
\EqRef{eq:beta2} for every pair of photons,
\begin{equation}
  \Coh_2=\Coh_1+\sum_{j<l}\Bigl[\prod_{i\ne j,l}\Sins_i\Bigr]\bet{2}(k_j,k_l),
  \label{eq:A2}
\end{equation}
which is \EqRef{eq:AJ} at $J=2$, and the real-virtual with the
helicity-averaged factor $\bar{\Virtual}$ of \EqRef{eq:delta-bar}, less the
resonant logarithm of \EqRef{eq:resonant-log}.
Together they give the full $\mathcal O(\alpha^2)$ square, the
double real and the real-virtual combined, replacing \EqRef{eq:A2} once the
loop is included,
\bea
  \avsq{\Coh_2^V}&=&\avsq{\Coh_2+\bar{\Virtual}\,\Coh_1}
  +\sum_{j\in\mathcal{R}}\Delta_j\,\Re\langle\Coh_2,\Coh_1^{(j)}\rangle \nonumber\\
  &+&2\,\Re\bigl(\Virtual^{(2)}\bigr)\avsq{\Coh_0} ,
  \label{eq:A2V}
\eea
where $\Coh_1^{(j)}=\bigl[\prod_{i\ne j}\Sins_i\bigr]\Mn{1}(k_j)$ is the
one-photon amplitude of photon $j$ summed over the assignments of the others,
and $\Virtual^{(2)}$ is the two-loop factor on the Born.
Unlike every other term here, $\Virtual^{(2)}$ is not generated
automatically from the tree amplitudes and the one-loop provider. It is
process-specific and has to be supplied from an external two-loop
calculation.
As at $\mathcal O(\alpha)$, the virtual multiplies the whole $\mathcal
  O(\alpha)$ amplitude $\Coh_1$ rather than $\Coh_0$. Its product with
$\bet{2}$ is of $\mathcal O(\alpha^3)$ and is dropped. The first term carries
the factorizable part, the virtual of the Born times the real, the second
the remainder of the emission process's own loop, and the third the double
virtual.
The helicity-averaged factor $\bar{\Virtual}$ in the first term is
the loop correction to the Born kinematics, not to the kinematics of emitting
photon $j$ itself, and the two agree only in the strict soft limit.
Multiplying $\Coh_1$ by $\bar{\Virtual}$ alone would therefore miss the
genuine dependence of photon $j$'s own loop on hard-photon kinematics. With
$\Virtual_j$ the IR-subtracted one-loop factor of the emission of photon $j$,
\beq
\Mn{1}(k_j)(1+\Virtual_j)=\Mn{1}(k_j)(1+\Virtual)+\Mn{1}(k_j)(\Virtual_j-\Virtual),
\eeq
and
$\Delta_j$ is the squared-level form $2\,\Re(\Virtual_j-\Virtual)$ of the bracket, so
that it enters the amplitude as $\tfrac12\Delta_j\,\Coh_1^{(j)}$ and the
square as \EqRef{eq:A2V}, up to terms of $\mathcal O(\alpha^3)$.
By the soft-photon theorem at one loop,
$\Mn{1}^{(1)}(k)\to \sfac(k)\,\Mzero^{(1)}$ as $k\to0$, and the subtraction
$\mathcal B_{\mathrm V}$ of the emission process tends to that of the Born,
since the photon is neutral and the charged legs only recoil. With the
external-photon counterterm, $\Delta_j\to0$ as $k_j\to0$. The soft-enhanced
real-virtual is therefore entirely in the first term, and the second is
next-to-soft.

\label{sec:loop-residuals}

The real-virtual is the first of a tower of one-loop residuals.
The residuals are formed automatically from the one-loop amplitudes the
provider returns. $\hat\beta_2^{[1]}$ is a run-time option, with the stability
checks of the real-virtual applied to each photon, and is not used for the
results of this paper. Higher multiplicities need only the corresponding
one-loop amplitudes.

\subsection{Weight Distribution}\label{sec:weight}

We write $\avsq{\Coh_k}=\hsum_{\hel}|\Coh_k(\hel)|^2$ for the spin-averaged
squares of the coherent amplitudes. The weight is formed against the density
with which the generator produced the event,
\begin{equation}
  \begin{split}
    \avsq{\Coh_{\mathrm{approx}}} ={} & \sum_\prt
    \prod_{j}\bigl|\sfac_{\prt(j)}(k_j)\bigr|^2
    \\       & \times
            \frac{s}{s_\prt}\,
            \hsum_{\hel}\bigl|\tilde\Mzero(\hel;s_\prt)\bigr|^2 ,
  \end{split}
  \label{eq:crude}
\end{equation}
where $\tilde\Mzero(\hel;s_\prt)$ is the on-shell Born amplitude at total momentum
$P-K^\prt_{\mathrm I}$, at the point at which the generator evaluates its own
Born.
This is the incoherent sum over the photon labels that the generator assigns
where each photon is generated from one dipole, with the eikonal on the
post-emission legs times the pair flux. The generator's event weight is
multiplied by
\begin{equation}
  \mathcal{W}_{\ACRAIC} = \frac{\avsq{\Coh_J^V}}{\avsq{\Coh_{\mathrm{approx}}}},
  \label{eq:w-acraic}
\end{equation}
with $\Coh_J^V$ the $\mathcal O(\alpha^J)$ amplitude including all virtual
corrections. The denominator is a sum of positive terms. The
coherent $\avsq{\Coh_0}$, by contrast, can be made arbitrarily small by cancellations
between partitions and is not a suitable normalizer.

At $\mathcal O(\alpha)$ the \ACRAIC weight \EqRef{eq:w-acraic} cannot be
negative. Its numerator is a sum over helicities of squared moduli, its
denominator \EqRef{eq:crude} a sum of positive terms, and the form factor
multiplies it as an exponential. This holds event by event, for every final
state, energy and photon multiplicity, and it survives the virtual correction
in both forms \EqRefs{eq:virt}{eq:ext-virtual}, since both enter inside the
modulus. At $\mathcal O(\alpha^2)$ the second term of \EqRef{eq:A2V} stands
outside the modulus and has no definite sign, so the $\mathcal O(\alpha^2)$
weight is not positive by construction. That term is next-to-soft and of
$\mathcal O(\alpha^2)$, so it can turn the weight negative only where the
weight is already small. The squared-level weight has no such property. In its
additive form, $1+v+\sum_j\Virtual_j$, each $\Virtual_j$ is negative wherever
the exact matrix element lies below the eikonal approximation. A factor
$1+\Virtual_j$ of the \YFSNLOEW approach is negative where the eikonal
approximation at the reduced kinematics exceeds the exact matrix element there
by more than the eikonal approximation at the event. \TabRef{tab:sign} counts the negative
weights for $e^+e^-\to\mu^+\mu^-$ at 240~\UGeV at $\mathcal O(\alpha)$. About
$0.3\%$ of the \YFSNLOEW weights are negative, a fraction that grows with the
photon multiplicity, and none of the \ACRAIC ones.

\begin{table}
  \centering\footnotesize
  \begin{tabular}{@{}lcc@{}}
    \toprule
    events         & $\YFSNLOEW$ & $\ACRAIC$ \\ \midrule all & $0.29\%$ & $0$ \\ ISR
    photons only   & $0.33\%$    & $0$       \\ with FSR photons & $0.27\%$ & $0$ \\ hardest
    FSR $x>0.05$   & $0.16\%$    & $0$       \\ hardest FSR $x<0.01$ & $0.32\%$ & $0$ \\
    $\le2$ photons & $0.10\%$    & $0$       \\ $\ge5$ photons & $0.45\%$ & $0$ \\
    \bottomrule
  \end{tabular}
  \caption{Fraction of negative event weights for $e^+e^-\to\mu^+\mu^-$ at
    240~\UGeV, $66<m_{\mu\mu}<116\,\UGeV$, at $\mathcal O(\alpha)$.}
  \label{tab:sign}
\end{table}
For a fraction $f$ of negative weights of comparable size, the effective number
of events $(\sum_iw_i)^2/\sum_iw_i^2$ is $(1-2f)^2$ of the generated number. At
the fractions of \TabRef{tab:sign} the loss is about one per cent, small for
this process, but the fraction grows with the photon and process multiplicity
and negative weights survive unweighting as events of weight $-1$. Those events
pass through detector simulation and reconstruction like any other, and
unfolding, reweighting and classifier training have to carry them explicitly.
Conversely, a negative \ACRAIC weight at $\mathcal O(\alpha)$ identifies a
defect. Positivity is not correctness, which the comparisons of
\SecRef{sec:validation} test, but it means that the precision required with
samples of $10^{12}$ $Z$ decays need not be bought with events that cancel
against each other.
Reducing negative weights and raising the unweighting efficiency is an active
field in event generation for hadron
colliders~\cite{Nachman:2020fff,Andersen:2021mvw,Andersen:2023cku,Andersen:2020sjs,Danziger:2021xvr,Danziger:2021eeg,Frederix:2020trv,Frederix:2023hom},
where the negative weights arise in the matching of next-to-leading-order
calculations to parton showers.
\ACRAIC does not produce negative weights at $\mathcal O(\alpha)$, so its
positive weights can be unweighted with any of the efficiency improvements
above, without the added complication of negative-weight events.

\section{Realization within \Sherpa}\label{sec:realisation}
The algorithm of \SecRef{sec:algorithm} needs, for every event, the tree
amplitudes, resolved in all external helicities, at the physical legs and with
partition-dependent propagator momenta, together with polarization vectors and
one-loop amplitudes that are consistent with them. This section states what
that requires of an automated amplitude generator, and how \Comix and the
one-loop providers interfaced to \Sherpa meet it.

\label{sec:bg}

A \BG recursion~\cite{Berends:1987me} builds an amplitude from off-shell
currents. The current of a set $\pi$ of external legs is assembled from the
currents of its subsets, \bea J(\pi) &=&
P(p_\pi)\sum_{N=2}^{n}\;\sum_{\substack{\{\pi_1,\dots,\pi_N\}\\
    \in\,\prt_N(\pi)}} \nonumber\\ && \hspace{2em}
V_N(\pi_1,\dots,\pi_N)\,J(\pi_1)\cdots J(\pi_N), \label{eq:bg-recursion}\\
p_\pi &=& \sum_{i\in\pi}p_i , \nonumber \eea where $\prt_N(\pi)$ is the set of
splittings of $\pi$ into $N$ disjoint non-empty subsets, $n$ is the largest
vertex multiplicity, $V_N$ is the $N$-point vertex, and $P(p_\pi)$ is the
propagator of the internal line with momentum $p_\pi$. The recursion starts
from the external wave functions $J(\{i\})$, and the amplitude is the current
of all legs but one contracted with the wave function of the last.
\Comix~\cite{Gleisberg:2008fv} implements \EqRef{eq:bg-recursion}, and its cost
for a fixed helicity configuration grows polynomially in the number of legs,
which makes the $(N+3)$-leg amplitude of \EqRef{eq:A1} affordable at
multiplicities where a diagrammatic expansion would not be. The remainders map
onto the recursion without modification. $\Mzero$ is the Born process, and
$\Mn{1}$ is the same process with the photon as one more external vector
current, requiring no new vertex and no hand-coded spinor algebra. Both are
returned for all helicity configurations in one evaluation. The
process-dependent content of \EqRef{eq:stage-partition-sum} is therefore
exactly the output of a general tree-level amplitude generator. It is also not
limited to the Standard Model, and the same recursion can be used for any model
that the generator supports through its UFO
interface~\cite{Degrande:2011ua,Darme:2023jdn,Hoche:2014kca}, including
effective field theories and BSM models. In particular, it applies to the QED
corrections in scalar QED, which is the standard description of photon
radiation off pions in
$e^+e^-\to\pi^+\pi^-(N\gamma)$~\cite{Hoefer:2001mx,Czyz:2003ue,Arbuzov:1997pj,Arbuzov:2005pt,Campanario:2019mjh,Colangelo:2022lzg,Ignatov:2022iou,Budassi:2024whw,Aliberti:2024fpq}.
\Sherpa's photon emission already treats scalar
emitters~\cite{Schonherr:2008av,Price:2025fiu}. A detailed study of this
process is the topic of a separate paper~\cite{Price:2026pion}.
The same applies to $\mu^\pm e^-\to\mu^\pm e^-$ at the MUonE
experiment, where the \YFS framework of \Sherpa has been applied at the
squared level, matched to the complete $\mathcal O(\alpha)$ and the dominant
$\mathcal O(\alpha^2)$ corrections~\cite{Price:2025pru}. The complete
two-loop virtual is available for a massless electron and a massive
muon~\cite{Bonciani:2021okt,Broggio:2022htr}, and the electron mass can be
restored by the massification used here, so it could replace the dominant
terms.

Because every current of \EqRef{eq:bg-recursion} is labelled by its leg set,
the shift rule of \SecRef{sec:partitions} is realized by replacing, per
partition, $P(p_\pi)$ by $P(p_\pi+\Delta_\pi)$ for each current $\pi$. The
shift enters the propagator denominators only and the vertex numerators and the
external wave functions keep the physical momenta. This gives each attachment
line its own invariant, which a single rescaled $s$-channel invariant cannot.
The $W$ decay stages of \SecRef{sec:stages} use the current masks of the $W$
decay products, so that a photon on a decay stage shifts exactly the $W$ line
of that decay in every contribution that contains one, and nothing else. The
production stage uses the mask of the initial legs, which puts the $s$-channel
line at $P-K$ by momentum conservation.

The coherent sum \EqRef{eq:stage-partition-sum} is not invariant under a
redefinition of the phase of an individual term and only the modulus of a
single term is. The amplitudes $\Mzero$ and $\Mn{1}$ and the polarization
vectors in \EqRef{eq:stage-soft-factor} must therefore be built with the same
gauge reference vector and the same spinor phase conventions, or $\bet{1}$
acquires a remainder that does not vanish in the soft limit. Such a mismatch is
invisible in any squared or helicity-summed quantity, and for photons along the
beam axis, where the phases of different conventions coincide up to a sign.
\ACRAIC therefore builds the soft factors from the photon polarization vectors
of \Comix itself (\AppRef{app:conventions}). An implementation that combines
amplitudes from one source with soft factors from another must instead
determine a relative phase and normalization.
The choice of the reference vector itself does not matter. The
\Comix amplitudes satisfy the Ward identity for the photon, and the stage soft
factors are invariant under a shift of the polarization vector by a multiple of
$k$ because every stage is neutral (\EqRef{eq:neutrality}), so only the common
phase convention has to be fixed.

\label{sec:providers}

The one-loop virtual of both \ACRAIC and \YFSNLOEW is generally taken from
\Sherpa's native interface to one-loop providers, such as
\OpenLoops~\cite{Buccioni:2019sur} and
\Recola~\cite{Actis:2016mpe,Denner:2017wsf} or for some select processes
internally coded amplitudes. The helicity-resolved factor \EqRef{eq:delta-hel}
requires the complex one-loop amplitude per helicity, which \Recola{} computes
and stores inside the call that the squared-level virtual already makes, so
that reading it out costs nothing further. \OpenLoops returns squared
quantities, and with it \ACRAIC uses the helicity-averaged form
\EqRef{eq:ext-virtual}. The same holds for the real-virtual.

For one-loop corrections in \YFS generators it has generally been a requirement
that all masses are kept. This can lead to numerical issues that we would
like to avoid. For leptons whose mass is small
against every invariant of the process, the massive one-loop amplitude can be
obtained from the massless one up to power
corrections~\cite{Catani:2000ef,Penin:2005eh,Mitov:2006xs,Becher:2007cu,Engel:2018fsb},
\begin{equation}
  \Mamp^{(m)}(\{p\})=\prod_{i}Z_i^{1/2}\,\Mamp^{(0)}(\{\tilde p\})
  +\mathcal O\bigl(m^2/|s_{ij}|\bigr),
  \label{eq:massification}
\end{equation}
where the product runs over the external leptons and, at one loop,
\begin{equation}
  Z_i^{(1)}=\frac{\alpha}{4\pi}\,Q_i^2\left[\frac{2}{\epsilon^2}
    +\frac{2L_i+1}{\epsilon}+L_i^2+L_i+4+\zeta_2\right] ,
  \label{eq:Z1}
\end{equation}
with $L_i=\ln(\mu^2/m_i^2)$, in the normalization of
~\cite{Engel:2018fsb}.
The relation \EqRef{eq:massification} is the basis of the two-loop
calculations of
~\cite{Penin:2005eh,Becher:2007cu,Engel:2018fsb} and of the two-loop
virtual in \McMule~\cite{Aliberti:2024fpq}, which uses massive amplitudes at
one loop and for $e^+e^-\to e^+e^-\gamma$ it was verified at one loop against
\OpenLoops in~\cite{Engel:2021ccn}. Here the mass expansion is
applied to the amplitude of any loop provider, independently of the process
and helicity by helicity, inside the amplitude-level exponentiation.
The loop provider is called with the external
legs as massless, at the massless $\{\tilde p\}$ of the event's
phase-space point with the momenta in the rest frame of the incoming pair with their
directions kept and the three-momenta rescaled by one common factor, which
conserves momentum exactly.

The factor of \EqRef{eq:delta-hel} becomes
\begin{equation}
  \Virtual(\hel)=\frac{a_1^{(0)}(\hel)}{a_0^{(0)}(\hel)}
  +\frac12\sum_i Z_i^{(1)}-\tfrac12\,\mathcal B_{\mathrm V} ,
  \label{eq:delta-massified}
\end{equation}
and the same factor multiplies the tree in the real-virtual of
\SecRef{sec:oalpha2}. The trees themselves remain the fully massive ones from
\Comix, so that only the ratio of loop to tree is approximated and the soft and
collinear structure of the real emission is unchanged. The poles of
$Z_i^{(1)}$ replace the collinear poles of the massless loop by the
logarithms of the lepton mass, so that the virtual cancels against the massive
$\mathcal B_{\mathrm V}$ exactly as without the option. The lepton masses of
the loop provider are global, so closed loops of the massified species are
massless as well. Without the photon self-energy this does not arise. With it,
in a scheme where $\Delta\alpha$ is an input such as $G_\mu$, it is a power
correction in $m^2$ over the virtuality of the loop.

\begin{figure*}
  \centering
  \includegraphics[width=0.32\textwidth]{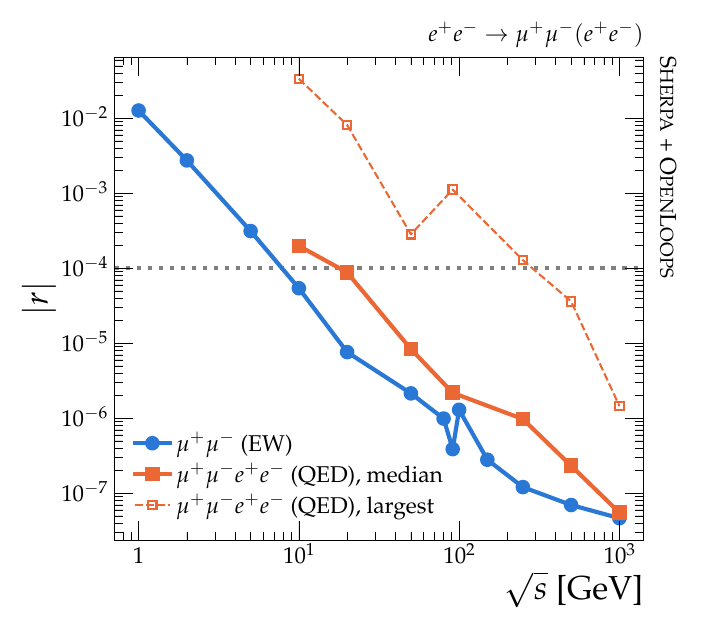}\hfill
  \includegraphics[width=0.32\textwidth]{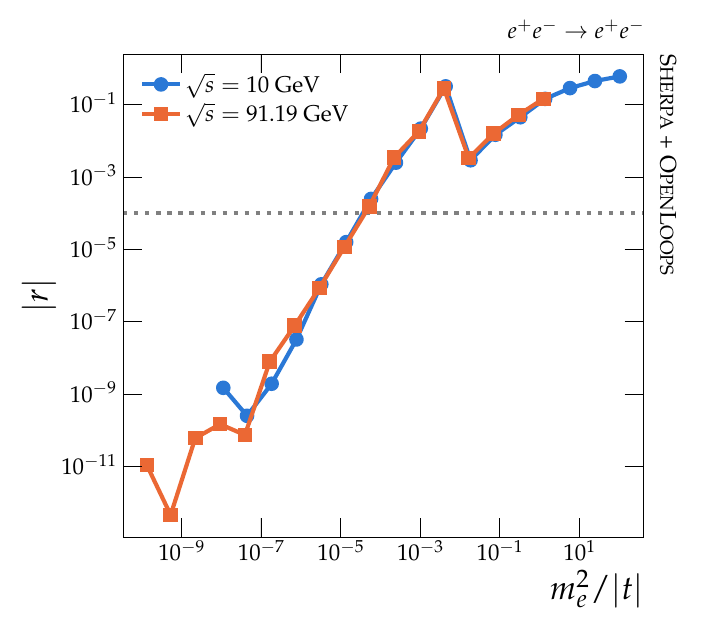}\hfill
  \includegraphics[width=0.32\textwidth]{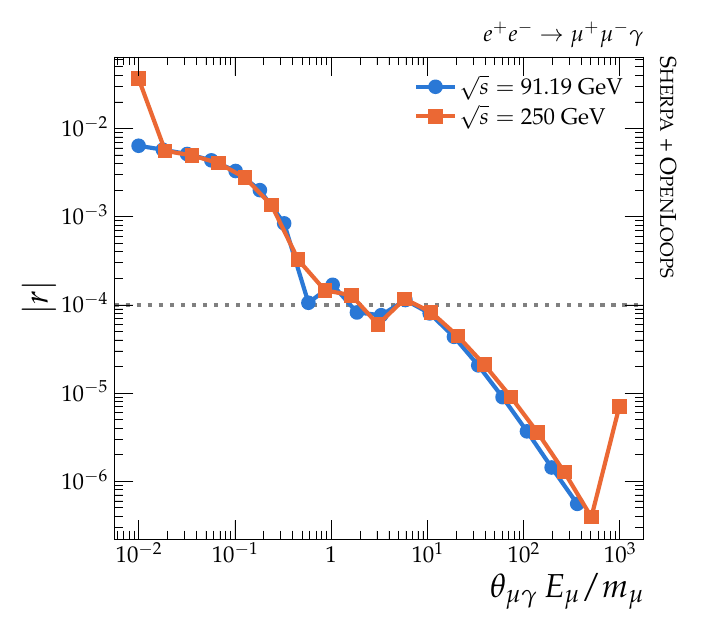}
  \caption{One-loop massification at fixed phase-space points, shown as the
    residual $r$ of \EqRef{eq:massification-residual} relative to the Born,
    from \OpenLoops. Top left: $e^+e^-\to\mu^+\mu^-$ in the full electroweak,
    the largest $|r|$ over five angles and
    two scales, and $e^+e^-\to\mu^+\mu^-e^+e^-$ with QED corrections, the
    median and largest $|r|$ over 50 random points per energy. Top right:
    Bhabha scattering against $m_e^2/|t|$. Bottom:
    $e^+e^-\to\mu^+\mu^-\gamma$ against the photon-muon angle in units of
    $m_\mu/E_\mu$.}
  \label{fig:massification}
\end{figure*}

\FigRef{fig:massification} shows the residual
\begin{equation}
  r=\Virtual^{(m)}-\Virtual^{(0)}-\sum_iZ_i^{(1)} ,
  \label{eq:massification-residual}
\end{equation}
with $\Virtual$ being the one-loop correction relative to the Born with and
without mass terms, at fixed phase-space points from \OpenLoops. For
$e^+e^-\to\mu^+\mu^-$ in the
full electroweak theory it is $\approx10^{-7}$ at $M_Z$, below
$3\times10^{-6}$ from 50~\UGeV upwards, and grows like $m_\mu^2/s$ towards
low energies. We expect the approximation to fail where the
power correction is not small. In the forward region of Bhabha scattering the
residual grows as $|t|$ decreases, reaching $3\times10^{-4}$ at $6.5~\mrad$
and staying below $10^{-4}$ for scattering angles above $13~\mrad$ at
$\sqrt s=M_Z$. In the real-virtual it is of order
$10^{-4}$ down to photon-lepton angles of $m/E$ and reaches a few $10^{-3}$
inside the dead cone, a region of small weight. Care should be taken at thresholds
where the Coulomb
term $\pi\alpha/(2\beta)$ becomes relevant and which the massless amplitude does not contain,
so that pairs produced near threshold, such as $\tau$ pairs at charm-factory
energies, require the massive amplitude. For the four-fermion final state
$e^+e^-\to\mu^+\mu^-e^+e^-$, with all six leptons massified and the QED
corrections only, the residual at random phase-space points is
$2\times10^{-6}$ (median) at $M_Z$ and is controlled by the largest
$m^2/|2p_i\cdot p_j|$ of the event, which bounds it at 99\% of the points.
The approximation therefore holds wherever every pair of leptons is resolved.

For photonic corrections, with no closed lepton loop, the factorization
\EqRef{eq:massification} holds at two loops with $Z_i=1+Z_i^{(1)}+Z_i^{(2)}$
and no soft function~\cite{Penin:2005eh,Becher:2007cu,Engel:2018fsb}. We use it
for the non-factorizable part of the photonic $\bet{0}^{(2)}$ of
$e^+e^-\to\ell^+\ell^-$, $\ell=\mu,\tau$, in \SecRef{sec:nnlo}. Applied to the
amplitude and to the initial- and final-state vertex form factors $F$, it gives
\bea \ln\Mamp^{(m)}-\ln F_{\rm ini}^{(m)}-\ln F_{\rm fin}^{(m)} &=&
\ln\Mamp^{(0)}-\ln F_{\rm ini}^{(0)}-\ln F_{\rm fin}^{(0)}\nonumber \\
&&+\mathcal O\bigl(m^2/|s_{ij}|\bigr), \label{eq:massification-2l} \eea since
every lepton enters the amplitude once and one form factor once, and its $Z_i$
cancel. The right-hand side is taken from the massless two-loop helicity
amplitudes of~\cite{Dave:2024ewl} and the massless form factor
of~\cite{Penin:2005eh}. Once the one-loop infrared function of the
initial-final dipoles is subtracted, its two-loop part has no infrared poles
and does not depend on the infrared scheme, and for massless leptons it is a
function of $t/s$ alone. It multiplies the photon-exchange part of each
helicity amplitude, while the massive vertex form factors on the left are
included exactly at two loops.
For example, in $e^+e^-\to\gamma\gamma$ the massless two-loop QED corrections are known in
closed form~\cite{Anastasiou:2002zn}. For this process, the photonic
$\bet{0}^{(2)}$ treats the $e^+e^-$ pair as a vertex. Once massified it is exact for the logarithms and the
massification constants, and differs from the complete result only in the hard
function, by an angle-dependent constant,
\bea
\Virtual^{(2)}\to\Virtual^{(2)}+\tfrac12\Bigl(\frac{\alpha}{2\pi}\Bigr)^2\Delta(t/s),\\
\Delta=\frac{2B_s}{T_0}-\frac{f_0}{2}-2r_F\,\Re\,r_1+r_F^2 .
\label{eq:diphoton-2l}
\eea
Here $B_s$ is the photonic two-loop interference
of~\cite{Anastasiou:2002zn}, $T_0=8(t/u+u/t)$ its tree-level normalization, and
$r_1$ the Catani-finite remainder of the one-loop $e^+e^-\to\gamma\gamma$
amplitude, the amplitude after subtracting its universal infrared poles in
Catani's one-loop factorization formula. $r_F=-4$ and $f_0$ are the
Catani-finite and real remainders,
respectively, of the massless one-loop vertex form factor, the same kind of
quantity as $r_1$ but for the vertex rather than the $e^+e^-\to\gamma\gamma$
amplitude, so that $\Delta$ vanishes for an $s$-channel vertex. It is
not applied where $m_e^2/\min(|t|,|u|)>10^{-3}$, outside the leading power. The
massified series also gives the infrared poles of the massive two-loop
amplitude, which must cancel against the YFS subtraction of the double
virtual,
\begin{equation}
  \Bigl(\tfrac12\mathcal B_{\mathrm V}^2+\mathcal B_{\mathrm V}V_1\Bigr)
  \avsq{\Mzero} ,
  \label{eq:double-virtual-sub}
\end{equation}
with $V_1$ the massive one-loop factor. Event by event, the
double and single poles agree to the size of $m_e^2/s$, and to better than
$10^{-6}$ for photons almost along a beam. This tests the two-loop
massification against the YFS form factor independently of the finite parts.

\subsection{Theory Uncertainties}\label{sec:uncertainty}

We estimate the missing higher orders with theory nuisance
parameters~\cite{Tackmann:2024kci,Lim:2024nsk} instead of the difference to the
next-lower order which is generally used for \YFS. The amplitudes of \EqRefs{eq:A1}{eq:A2} are sums of classes,
$\Coh=\sum_c\Coh_c$. They are the $\bet{0}$ term, the $\bet{1}$ of the photons
assigned to the initial and to the final state, the factorized multi-photon
remainder, and at $\mathcal O(\alpha^2)$ the $\bet{2}$ of initial-state,
final-state and mixed photon pairs. Each class gives a template $T_c$ on the
same events, the response of the event weight \EqRef{eq:w-acraic} to a
rescaling of $\Coh_c$ by $1+\epsilon$. Each nuisance parameter
$\theta_t\in[-1,1]$ multiplies one template with a normalization $N_t$, the
leading-logarithmic size of the first term the class omits, and the band is the
sum in quadrature of the $\theta_t=\pm1$ variations,
\begin{align}
  \mathrm d\sigma(\theta) & =\mathrm d\sigma\,\Big[1+\sum_t\theta_t N_t T_{c(t)}\Big] , \\
  T_c                     & =\frac{2\,\Re\langle\Coh_c,\Coh_J^V\rangle}{\avsq{\Coh_J^V}} ,
  \qquad \sum_c T_c=2 ,
  \label{eq:tnp-band}                                                                                 \\
  \delta\sigma            & =\Big[\sum_t\Big(N_t\!\int\!\mathrm d\sigma\,T_{c(t)}\Big)^2\Big]^{1/2} ,
  \label{eq:tnp-dsigma}                                                                               \\
  g_{e,f}                 & =\frac{\alpha}{\pi}\Big(\ln\frac{s}{m_{e,f}^2}-1\Big) ,
  \label{eq:tnp-g}                                                                                    \\
  N_{\bet{1}}             & =\frac12\,\frac{g^k}{k!} ,\qquad N_{\bet{2}}=\frac12\,g ,
  \label{eq:tnp-n1}                                                                                   \\
  N_{\bet{0}}^{\rm ini}   & =\frac12\,\frac{g_e^m}{m!} ,\qquad
  N_{\bet{0}}^{\rm fin}=\frac12\,\frac{Q_f^2\,g_f^m}{m!} ,
  \label{eq:tnp-n0}                                                                                   \\
  N_{\bet{0}}^{\times}    & =\frac{(g_e+Q_f^2g_f)^m-g_e^m-(Q_f^2g_f)^m}{2\,m!} .
  \label{eq:tnp-n0x}
\end{align}
Here $g$ is $g_e$ for the initial-state and multi-photon classes and $g_f$ for
the final-state ones, $k$ is the order at which the run is truncated, and $m$ is
the first order of $\bet{0}$ missing from the prediction, $m=2$ at
$\mathcal O(\alpha)$ and $m=3$ at $\mathcal O(\alpha^2)$ with the photonic
$\bet{0}^{(2)}$ included in the prediction. The three $\bet{0}$ parameters act on
the same template, and $N_{\bet{0}}^{\times}$ collects the products of initial-
and final-state logarithms of the same order. The templates are
constructed on the fly using \Sherpa's multi-weight infrastructure~\cite{Bothmann:2016nao}.

For the missing electroweak corrections at the $Z$ pole we evaluate the
two-loop electroweak corrections of \Griffin~\cite{Chen:2022dow}, the step from
the complete one-loop corrections in the $G_\mu$ scheme of our inputs, on the
same events. Where the prediction does not include this step,
  its size is the electroweak uncertainty. Where it does, the uncertainty is the
  error estimate built into \Griffin. It gives the missing three-loop and mixed
  electroweak--QCD terms of~\cite{Dubovyk:2019szj} as errors on the axial-vector
  form factor and on the effective mixing angle of each fermion. We vary both by
  these errors in the \Griffin matrix element itself, with $\Gamma_Z$ held at
  its input value as in the prediction, take the errors of the initial and final
  leptons as fully correlated, and add the two variations in quadrature. For
  $e^+e^-\to\mu^+\mu^-$ and $e^+e^-\to\tau^+\tau^-$ at the peak this gives
  $2.4\times10^{-4}$ of the cross section, and the error on
  $\sin^2\theta_{\rm eff}^\ell$, $4.3\times10^{-5}$, gives $7.6\times10^{-5}$ in
  the forward-backward asymmetry. We take the estimate from \Griffin rather than
  propagating the pseudo-observable errors of~\cite{Dubovyk:2019szj} by hand.
  It is then independent of our own analysis, and it follows any improved
  estimate in a later release of \Griffin without further work on our side.
  Both errors lie below the $\mathcal O(\alpha)$ photonic band. They are errors on the $Z$-pole
  pseudo-observables and so cover the resonant part of the matrix element, which
  \Griffin evaluates at next-to-next-to-leading order. The non-resonant photon
  exchange and its interference with the $Z$ are included only at next-to-leading
  order~\cite{Chen:2022dow}, and their missing higher orders, which matter for
  events whose invariant mass after initial-state radiation lies below the peak,
  are not in the band.
We also
give the difference between the $\alpha(M_Z)$ and $G_\mu$ input schemes on the
same events, which probes the input-scheme dependence of the one-loop
prediction. It stands for the known next-to-next-to-leading-order corrections to
$e^+e^-\to ZH$, none of which is in our prediction. The mixed electroweak--QCD
corrections are about $1\%$ of the Born cross
section~\cite{Gong:2016jys,Sun:2016bel} and $1.5\%$ for $\mu^+\mu^-H$ with the
finite $Z$ width~\cite{Chen:2018xau} in the $\alpha(0)$ scheme, but $0.3\%$ in
the $G_\mu$ scheme of our inputs~\cite{Sun:2016bel}. The two-loop electroweak
corrections with closed fermion loops are $+0.7\%$ at 240~\UGeV in the
$\alpha(0)$ scheme~\cite{Freitas:2022hyp}, and the complete two-loop electroweak
amplitudes are known only unrenormalized~\cite{Chen:2022mre}.

Light-fermion pair corrections are not generated by \ACRAIC and are in neither
band. For $e^+e^-\to\mu^+\mu^-$ with $s'>0.5\,s$, close to our
$m_{\mu\mu}>60~\UGeV$, the initial-state pair corrections, summed over flavours
and including their higher orders, are about $-2.5\times10^{-3}$ at and below
the peak and vanish $3~\UGeV$ above it, with an uncertainty of a few
$10^{-4}$~\cite{Arbuzov:1999xso}. The final-state ones are a few $10^{-4}$. At
these energies soft and virtual pairs give the bulk of the correction, so the
hard real pairs of a tree-level $2\to4$ calculation do not estimate it. They
have to be added from the literature until \ACRAIC generates pairs.
The initial-state piece alone is an order of magnitude above the
photonic $\mathcal O(\alpha^2)$ band.

\section{Validation}\label{sec:validation}
In this section we validate our \ACRAIC implementation. We start with
conditions that hold on every event, namely the internal consistency conditions
of \SecRef{sec:algorithm}, the event-by-event agreement with \YFSNLOEW, and the
closure of the tree residuals against the exact multi-photon amplitudes. We
then compare with \kkmc, where an independent amplitude-level implementation
exists, in cross sections and distributions at photonic $\mathcal O(\alpha^2)$.
We then compare the fiducial cross sections with \YFSNLOEW across final states,
where the two share the resummation and differ only in the matching.
\SecRef{sec:nnlo} turns to the step to $\mathcal O(\alpha^2)$ and the
perturbative uncertainty, with $Z$-pole precision distributions as an example
application, and \SecRef{sec:recoil} applies the same machinery, as a further
example application rather than a validation in itself, to the Higgs recoil
mass at FCC-ee. \TabRef{tab:summary} lists the processes, the order of the
\ACRAIC prediction for each and the source of its two-loop content. Every cross
section and distribution in this section is computed with
\Rivet~\cite{Bierlich:2024vqo}, storing the weight variations of
\SecRef{sec:uncertainty} with \YODA~\cite{Buckley:2023fyu}.

\begin{table*}
  \centering\small
  \begin{tabular}{@{}l l c l@{}}
    \toprule
    $\sqrt s$ [GeV] & process & \ACRAIC & two-loop content \\
    \midrule
    \multirow{6}{*}{91.19}
      & $\mu^+\mu^-$, $\tau^+\tau^-$
      & $\mathcal O(\alpha^2)$
      & \parbox[t]{0.55\textwidth}{\raggedright $\bet{0}^{(2)}$: logarithms, vertex constants~\cite{Penin:2005eh},
        non-factorizable boxes~\cite{Dave:2024ewl} by massification. Dyson-resummed
        vacuum polarization. \Griffin~\cite{Chen:2022dow} two-loop electroweak step in
        \FigRef{fig:tnp}}\\
      & $e^+e^-$
      & $\mathcal O(\alpha^2)$
      & \parbox[t]{0.55\textwidth}{\raggedright $\bet{0}^{(2)}$: logarithms, vertex constants of the $s$-channel vertices. Dyson-resummed
        vacuum polarization}\\
      & $\nu\bar\nu$
      & $\mathcal O(\alpha^2)$
      & \parbox[t]{0.55\textwidth}{\raggedright $\bet{0}^{(2)}$ of the initial state: logarithms, vertex constant. Dyson-resummed
        vacuum polarization}\\
      & $d\bar d$, $u\bar u$
      & $\mathcal O(\alpha^2)$
      & \parbox[t]{0.55\textwidth}{\raggedright $\bet{0}^{(2)}$: logarithms, vertex constants. Dyson-resummed vacuum polarization}\\
      & $\gamma\gamma$
      & $\mathcal O(\alpha^2)$
      & \parbox[t]{0.55\textwidth}{\raggedright $\bet{0}^{(2)}$ with the massless two-loop
        amplitude~\cite{Anastasiou:2002zn} and the two-loop massification
        constants~\cite{Mitov:2006xs}.}\\
      & $f\bar f\gamma$
      & $\mathcal O(\alpha)$
      & \parbox[t]{0.55\textwidth}{\raggedright --}\\
    \midrule
    161, 250
      & $\mu^+\mu^-$
      & $\mathcal O(\alpha^2)$
      & \parbox[t]{0.55\textwidth}{\raggedright as at the $Z$ pole, without the \Griffin step}\\
    \midrule
    \multirow{3}{*}{240}
      & $H\mu^+\mu^-$, $H\nu\bar\nu$
      & $\mathcal O(\alpha^2)$
      & \parbox[t]{0.55\textwidth}{\raggedright $\bet{0}^{(2)}$: logarithms, vertex constants. No two-loop electroweak}\\
      & four leptons, $\gamma\gamma$
      & $\mathcal O(\alpha)$
      & \parbox[t]{0.55\textwidth}{\raggedright --}\\
      & $H f\bar f\gamma$
      & $\mathcal O(\alpha)$
      & \parbox[t]{0.55\textwidth}{\raggedright --}\\
    \bottomrule
  \end{tabular}
  \caption{The order of the \ACRAIC prediction for each process and the source of
    its two-loop content. The logarithms of $\bet{0}^{(2)}$ are
    $\tfrac12\delta_f^2+\delta_f w+\delta_{\rm NLL}$ of \EqRef{eq:beta02-content}, the
    leading- and next-to-leading-logarithmic terms of the two-loop photonic
    vertex form factor, and the vertex constants those of the massive
    two-loop form factor~\cite{Penin:2005eh}. Every $\mathcal O(\alpha^2)$ prediction also
    contains the exact double real from \Comix and the real-virtual from the one-loop
    provider. Beyond $\mathcal O(\alpha^2)$ only the soft logarithms of the \YFS exponent
    are included.}
  \label{tab:summary}
\end{table*}

Our implementation in~\cref{sec:realisation} rests on statements that can be
checked on every event. At $\Nstage=2$ and $N=2$ the general construction
reproduces the dedicated $2\to2$ implementation to rounding, in the soft
factors, where the per-leg sum \EqRef{eq:stage-soft-factor} evaluates a neutral
pair in a different order, and in the density matrix. Neutrality,
\EqRef{eq:neutrality}, is imposed as a precondition on the stage decomposition,
and closure, \EqRef{eq:closure}, is checked numerically per photon. The number
of partitions is checked against $\Nstage^m$ on every event. The one-photon
closure, $\Coh_1=\Mn{1}(k)$, holds to rounding for $e^+e^-\to H\mu^+\mu^-$
($\Nstage=2$) and for $e^+e^-\to\mu^+\mu^-\tau^+\tau^-$ ($\Nstage=3$), as for
$2\to2$, and on two-photon $\mu^+\mu^-\tau^+\tau^-$ events $\bet{1}$ is bounded
by the Born in every helicity configuration.

\begin{figure*}
  \centering
  \includegraphics[width=0.32\textwidth]{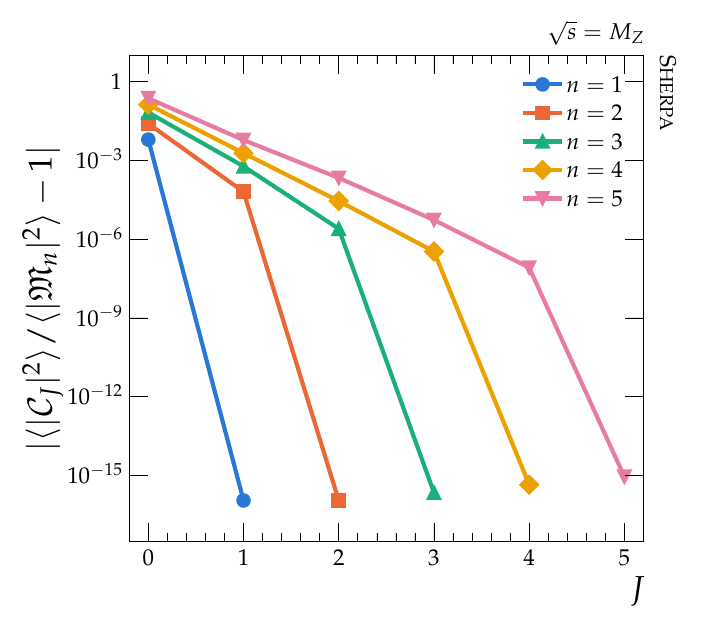}\hfill
  \includegraphics[width=0.32\textwidth]{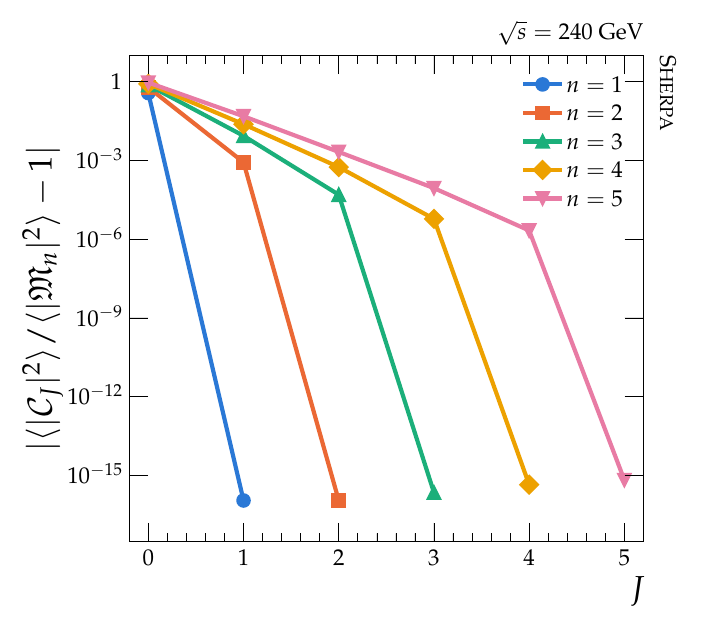}\hfill
  \includegraphics[width=0.32\textwidth]{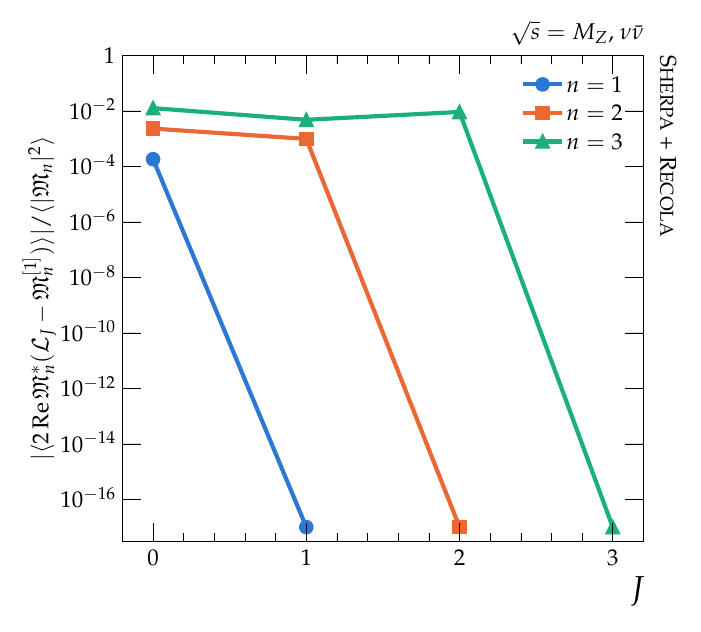}
  \caption{Truncation of the residuals. Left and middle: the partition- and
  helicity-summed $\avsq{\Coh_J}$ of \EqRef{eq:AJ} against the exact
  $n$-photon tree amplitude $\avsq{\Mn{n}}$ from \Comix, on
  $e^+e^-\to\mu^+\mu^-+n\gamma$ events at $\sqrt s=M_Z$ and
  \newinline{240}~\UGeV. Right: the one-loop residuals,
  the interference of their truncation $\mathcal L_J$ with the tree against
  that of the one-loop amplitude $\Mn{n}^{[1]}$, for
  $e^+e^-\to\nu_\mu\bar\nu_\mu+n\gamma$ at $\sqrt s=M_Z$.}
  \label{fig:betan}
\end{figure*}

Where the two matchings must coincide, they can be compared event by event,
since both columns are evaluated on the same events and multiply the same
generator weight. We compare the real-only correction factors for \ACRAIC and
the \YFSNLOEW factor divided by $(1+v)$. One difference in bookkeeping has to
be removed first. \ACRAIC samples one helicity per photon, so per event
$\avsq{\Mn{1}}$ enters for that helicity only, while \YFSNLOEW sums over it.
The two agree on average but differ point by point. Two runs with the photon
helicity fixed to $+1$ and $-1$ draw the same random numbers, so they produce
identical events, and their average is the helicity sum. With the virtual, the
two columns compose it differently with the Born. \YFSNLOEW multiplies the Born
by $1+v$, with $v=\betsq{0}^{1}/\betsq{0}^{0}$. \ACRAIC multiplies the
amplitude by $1+\Virtual$ and squares, which with the helicity-averaged
$\Virtual=\bar{\Virtual}=v/2$ of \EqRef{eq:delta-bar} gives, per event,
\begin{align}
  \frac{\hsum_{\hel}|(1+\Virtual)\Mzero|^2}{\hsum_{\hel}|\Mzero|^2}
   & =(1+v/2)^2=1+v+\frac{v^2}{4} ,
  \label{eq:vsq-ratio}              \\
  \frac{w_{\ACRAIC}}{w_{\YFSNLOEW}}
   & =1+\frac{v^2}{4(1+v)} ,
  \label{eq:vsq-excess}
\end{align}
the $\hsum_{\hel}|\bet{0}^{(1)}|^2$ term of \EqRef{eq:virtual-identity} that the
squared-level matching does not carry. The
term is of $\mathcal O(\alpha^2)$, and accounts for the largest single contribution to the
difference between the two matching schemes.
The remainders of \EqRef{eq:AJ} are tested point by point against the exact
tree amplitude. On events with $n\le5$ photons, all of them in every set, the
\bet{} of \EqRef{eq:AJ} are evaluated for $J=0,\dots,n$ and
compared with the $n$-photon amplitude from \Comix at the same momenta,
evaluated without propagator shifts or partitions. At $J=n$ the closure holds
to rounding at both energies and for every $n$, as shown in \FigRef{fig:betan}.
Below $J=n$ each order removes one power of the photon energy fractions.

The one-loop residuals close in the same way. On
$e^+e^-\to\nu_\mu\bar\nu_\mu+n\gamma$ events at the $Z$ pole with $n=1$, $2$
and $3$ real photons, the right panel of \FigRef{fig:betan} compares the
interference of the one-loop truncation $\mathcal L_J$ with the tree with that
of the infrared-subtracted one-loop amplitude $\Mn{n}^{[1]}$ at the same
momenta. At $J=n$ the two agree to rounding, so the subtraction holds on the
full one-loop amplitudes with two and three photons, which enter through
$\hat\beta_2^{[1]}$ and $\hat\beta_3^{[1]}$. In principle, this means that the
one-loop residuals can be computed for any multiplicity with the only
limitation being the availability of the one-loop provider.

\begin{figure*}
  \centering
  \includegraphics[width=0.72\textwidth]{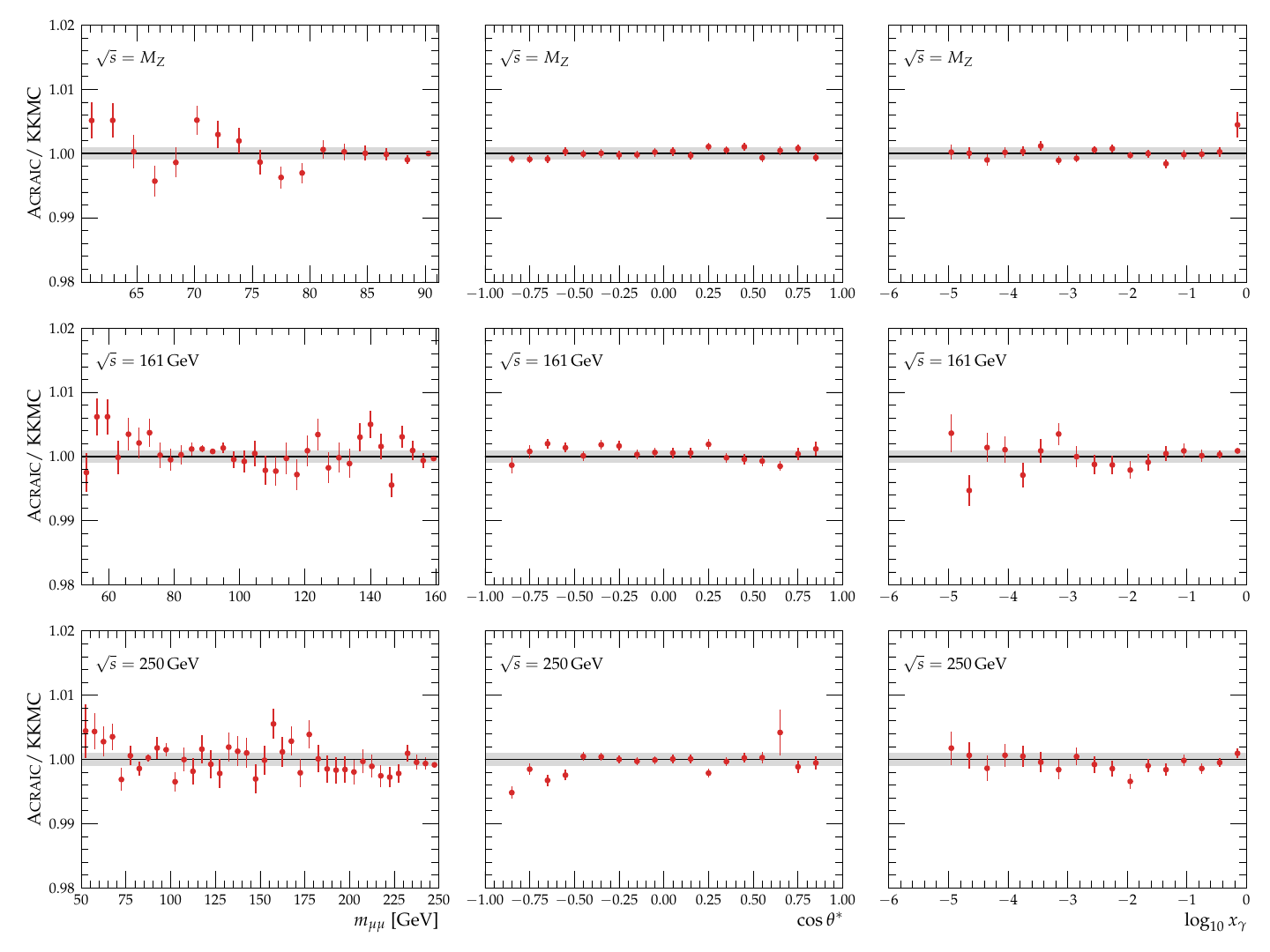}
  \caption{\ACRAIC at photonic $\mathcal O(\alpha^2)$, with $\bet{0}^{(2)}$, over \kkmc
    \OCEEX{2} for $e^+e^-\to\mu^+\mu^-$, at $\sqrt s=M_Z$ (top), $161$ (middle)
    and $250\,\UGeV$ (bottom). Left: the pair mass. Centre: $\cos\theta^*$, the polar angle in
    the pair rest frame. Right: the energy fraction of the hardest photon. The
    error bars are the combined statistical errors, the grey band marks
    $\pm10^{-3}$.}
  \label{fig:kkmc}
\end{figure*}

In \FigRef{fig:kkmc} we compare \ACRAIC at photonic $\mathcal
  O(\alpha^2)$, with $\bet{0}^{(2)}$, with \kkmc's \OCEEX{2}, both with
initial--final interference, at the $Z$ pole, $161$ and $250\,\UGeV$. The two
calculations are not identical at this order. \ACRAIC contains the complete
real-virtual contributions, the one-loop corrections to the emission of a hard
photon, that \OCEEX{2} does not include. The fiducial cross sections agree
within $0.5\times10^{-3}$ at all three energies. The differences appear where
the real-virtual contributions act. With a tagged hard photon the cross section
of \ACRAIC is higher by $2.6(6)\times10^{-3}$ at $161\,\UGeV$ and by
$2.5(5)\times10^{-3}$ at $250\,\UGeV$, while at the $Z$ pole the two agree
within $0.5\times10^{-3}$. At $250\,\UGeV$ the asymmetries differ by
$0.6(2)\times10^{-3}$. Apart from these, the distributions agree within the
statistical errors, including the radiative tail below the $Z$ peak.

\TabRef{tab:fid} compares \ACRAIC with \YFSNLOEW for thirteen processes at the $Z$
pole and at 240~\UGeV. The hard process is evaluated in the $G_\mu$ input scheme
and photons are emitted with $\alpha(0)$, with the input parameters of
\TabRef{tab:ew-params} in the complex-mass scheme~\cite{Denner:1991kt,Denner:2005fg}.
\begin{table}
  \centering\small
  \begin{tabular}{@{}l l@{\hspace{2em}}l l@{}}
    \toprule
    $G_F$    & $1.16639\times10^{-5}\,\UGeV^{-2}$ & $1/\alpha(0)$ & $137.03599976$  \\
    $M_Z$    & $91.1876\,\UGeV$                   & $\Gamma_Z$    & $2.4952\,\UGeV$ \\
    $M_W$    & $80.379\,\UGeV$                    & $\Gamma_W$    & $2.085\,\UGeV$  \\
    $M_H$    & $125.09\,\UGeV$                    & $\Gamma_H$    & $4.1\,\UMeV$    \\
    $m_t$    & $172.5\,\UGeV$                     & $\Gamma_t$    & $1.32\,\UGeV$   \\
    $m_e$    & $0.511\,\UMeV$                     & $m_\mu$       & $105\,\UMeV$    \\
    $m_\tau$ & $1.777\,\UGeV$                     & $m_u,\ m_d$   & $5,\ 10\,\UMeV$ \\
    \midrule
    \multicolumn{4}{@{}l@{}}{derived: $1/\alpha_{G_\mu}=132.119$,\quad
    $\sin^2\theta_W=0.223043-0.001105\,i$}                                          \\
    \bottomrule
  \end{tabular}
  \caption{Electroweak input parameters, \Sherpa's defaults. $\Gamma_\tau=0$, and
    $\Gamma_H=0$ where the Higgs boson is external. The quark masses enter only
    the $u\bar u$ and $d\bar d$ final states, as regulators of the collinear
    photon emission.}
  \label{tab:ew-params}
\end{table} Both columns carry the same complete
electroweak one-loop virtual. The cross sections are fiducial, using generic cuts applied to final-state particles,

\begin{align}
   & |\cos\theta|<0.9                                     &  & \text{visible pair members,} \nonumber                  \\
   & 60\, \UGeV <m_{f\bar f}                              &  & Z\ \text{pole,} \nonumber                               \\
   & 66\, \UGeV <m_{\ell\ell}<116\,\UGeV                  &  & \text{lepton pairs, } \newinline{240}\,\UGeV, \nonumber \\
   & E_\gamma>0.3\, \sqrt s,\ |\cos\vartheta_\gamma|<0.85 &  &
  \label{eq:fid-cuts}
\end{align}

The angle $\theta^*$ in the distributions is the polar angle of the fermion in
the $f\bar f$ rest frame, measured from the bisector of the electron direction
and the reversed positron direction in that frame, as in
Ref.~\cite{Collins:1977iv}. Without initial-state radiation it is the angle to
the electron beam. \YFSLO is the Born-level of the same resummation, without
$\mathcal O(\alpha)$ matching, it is not the tree-level Born cross section.
\ACRAIC exponentiates the resonant initial-final logarithm into the propagators
and puts the virtual on the whole $\mathcal O(\alpha)$ amplitude.

\begin{table*}
  \centering\small
  \sisetup{table-align-uncertainty=true}
  \begin{tabular}{@{}l l c c c S[table-format=1.7(2)]@{}}
    \toprule
    & & \multicolumn{3}{c}{$\sigma_\mathrm{fid}$ [pb]} & \\
    \cmidrule(lr){3-5}
    {$\sqrt s$ [GeV]} & {final state} & {\YFSLO} & {\YFSNLOEW} & {\ACRAIC} & {\ACRAIC/\YFSNLOEW} \\
    \midrule
    \multirow{7}{*}{91.19} & $\mu^+\mu^-$ & 1178.94(16) & 1234.41(18) & 1239.08(19) & 1.003782(16) \\
     & $\tau^+\tau^-$ & 1187.20(19) & 1242.61(21) & 1246.92(22) & 1.003466(18) \\
     & $e^+e^-$ & 1682.7(5) & 1777.4(6) & 1784.3(6) & 1.00391(6) \\
     & $\nu\bar\nu$ & 2747.55(27) & 2909.68(29) & 2915.11(29) & 1.0018653(10) \\
     & $d\bar d$ & 5261.8(13) & 5500.2(13) & 5513.0(14) & 1.002332(14) \\
     & $u\bar u$\ & 4072.4(4) & 4241.3(4) & 4244.9(4) & 1.00086(14) \\
     & $\gamma\gamma$ & 28.6302(27) & 27.6471(27) & 27.6342(27) & 0.99953(14) \\
    \midrule
    & & \multicolumn{3}{c}{$\sigma_\mathrm{fid}$ [fb]} & \\
    \cmidrule(lr){3-5}
    {$\sqrt s$ [GeV]} & {final state} & {\YFSLO} & {\YFSNLOEW} & {\ACRAIC} & {\ACRAIC/\YFSNLOEW} \\
    \midrule
    \multirow{7}{*}{240} & $H\mu^+\mu^-$ & 4.640(8) & 4.723(9) & 4.732(9) & 1.00192(25) \\
     & $H\nu\bar\nu$ & 12.596(6) & 12.881(6) & 12.912(6) & 1.002397(15) \\
     & $\mu^+\mu^-\tau^+\tau^-$ & 1.6776(15) & 1.773(8) & 1.750(4) & 0.987(5) \\
     & $\tau^+\tau^-\tau^+\tau^-$ & 0.9592(13) & 1.029(6) & 1.056(6) & 1.026(8) \\
     & $\tau^+\tau^-\nu_\mu\bar\nu_\mu$ & 3.5058(11) & 3.597(7) & 3.610(13) & 1.004(4) \\
     & $\gamma\gamma$ & 4133.2(4) & 3976.9(4) & 3974.3(4) & 0.99936(15) \\
    \midrule
    & & \multicolumn{3}{c}{$\sigma_\mathrm{fid}$ [pb]} & \\
    \cmidrule(lr){3-5}
    {$\sqrt s$ [GeV]} & {final state} & {\YFSLO} & {\YFSNLOEW} & {\ACRAIC} & {\ACRAIC/\YFSNLOEW} \\
    \midrule
    365 & $t\bar t$\,\textsuperscript{\dag} & 0.37437(18) & 0.41554(20) & 0.41689(21) & 1.003251(4) \\
    \bottomrule
  \end{tabular}
  \caption{Fiducial cross sections at $\mathcal O(\alpha)$ with the inputs of \TabRef{tab:ew-params}. Stable top quarks, no fiducial cuts ($\sigma_\mathrm{fid}=\sigma_\mathrm{tot}$).}
  \label{tab:fid}
\end{table*}

At the $Z$ pole \ACRAIC and \YFSNLOEW agree to $0.4\%$, \ACRAIC being higher by
$0.35$--$0.4\%$ for the charged leptons and by less for the other final states,
except for $\gamma\gamma$, where it is lower by $0.05\%$. At 240~\UGeV they
agree to $0.25\%$ for $\gamma\gamma$, $H\nu\bar\nu$ and $H\mu^+\mu^-$, and
within one standard deviation for $\tau^+\tau^-\nu_\mu\bar\nu_\mu$. For
$\mu^+\mu^-\tau^+\tau^-$ \ACRAIC is lower by $1.3(5)\%$ and for
$\tau^+\tau^-\tau^+\tau^-$ higher by $2.6(8)\%$, two and three standard
deviations. As in \SecRef{sec:rad}, a few large weights dominate the \YFSNLOEW
column of these four-lepton final states, which we attribute to poor numerical
convergence rather than a defect of either prediction.
$\mu^+\mu^-\nu\bar\nu$ contains $W$-pair diagrams, which the resonant
exponentiation of \SecRef{sec:virtual} does not yet cover.

The $Z$-pole ratios of \TabRef{tab:fid} differ from one by up to
$4\times10^{-3}$, and the difference can be accounted for term by term. For the
charged final states at the $Z$ pole we rerun \ACRAIC with and without the
exponentiated resonant logarithm on the same events as \TabRef{tab:fid}, and
split the difference into three pieces (\TabRef{tab:budget}),
\begin{equation}
  T\equiv\frac{\sigma_{\ACRAIC}}{\sigma_{\YFSNLOEW}}-1=\Gres+V+R .
  \label{eq:budget-split}
\end{equation}
$\Gres$ is the resonant initial-final logarithm of
\EqRef{eq:resonant-log} exponentiated into the propagators, the ratio of the
two \ACRAIC runs. $V$ is the square of the virtual, the term $v^2/4$ of
\EqRef{eq:vsq-excess}, which the squared-level matching does not carry. $R$ is
everything else, evaluated per event as the \ACRAIC factor without $\Gres$ minus
the \YFSNLOEW factor minus $V$. $\Gres$ and $V$ are the only terms the two
matchings define differently, and both are of $\mathcal O(\alpha^2)$, so $R$
tests the $\mathcal O(\alpha)$ content they share. At the $Z$ pole $R$ is below
$10^{-4}$ for every final state except $e^+e^-$, where it is $-1.8(5)\times10^{-4}$,
and the two columns agree to $2\times10^{-4}$ once $\Gres$ and $V$ are removed, while
$T$ is up to twenty times larger. The fiducial
differences of \TabRef{tab:fid} therefore measure how far two matchings that
coincide at $\mathcal O(\alpha)$ differ at $\mathcal O(\alpha^2)$, not an error
of either. $V$ is fixed only by the two-loop virtual, and its size follows that
of $v$ in the input scheme, at the $Z$ pole about $0.1$ in the $G_\mu$ scheme.

\begin{table}
  \centering\footnotesize
  \begin{tabular}{@{}lcccc@{}}
    \toprule
    final state    & $T$         & $\Gres$    & $V$    & $R$         \\
    \midrule
    $\mu^+\mu^-$   & $37.8(2.1)$ & $10.2$ & $27.2$ & $-0.1(0.4)$ \\
    $\tau^+\tau^-$ & $34.7(2.5)$ & $10.5$ & $23.1$ & $0.8(0.4)$  \\
    $e^+e^-$       & $39.1(4.7)$ & $4.2$  & $35.8$ & $-1.8(0.5)$ \\
    $d\bar d$      & $23.3(3.5)$ & $4.1$  & $18.0$ & $0.9(0.1)$  \\
    $u\bar u$      & $8.6(1.4)$  & $-3.4$ & $11.4$ & $0.5(0.1)$  \\
    \bottomrule
  \end{tabular}
  \caption{The difference $T$ of \EqRef{eq:budget-split} of the fiducial
    cross sections of \TabRef{tab:fid}, split into
    the exponentiated resonant logarithm ($\Gres$), the square of the virtual ($V$)
    and the rest ($R$), in units of $10^{-4}$, with the cross term
    $\Gres(V+R)<0.1\times10^{-4}$.}
  \label{tab:budget}
\end{table}

The \ACRAIC $\mathcal O(\alpha)$ column is not a strict truncation, since it
carries the term $(v/2)^2$ and the virtual on the whole $\mathcal O(\alpha)$
amplitude, both of $\mathcal O(\alpha^2)$. \TabRef{tab:fid-strict} gives the
strict truncation, with the virtual on $\Coh_0$ only and no $(v/2)^2$, on the
same events. Removing the two terms lowers the \ACRAIC cross sections at the
$Z$ pole by $0.3$--$1.6\times10^{-3}$ and that of $H\nu\bar\nu$ at 240~\UGeV by
$2.5\times10^{-3}$. The strict column then agrees with \YFSNLOEW within
$2.5\times10^{-4}$ for $\nu\bar\nu$, $u\bar u$ and $H\nu\bar\nu$. For the
charged leptons, $d\bar d$, $\gamma\gamma$ and $H\mu^+\mu^-$ a difference of
$1$--$4\times10^{-3}$ remains, which includes the exponentiated resonant
logarithm $\Gres$ of \TabRef{tab:budget}.

\begin{table*}
  \centering\small
  \sisetup{table-align-uncertainty=true}
  \begin{tabular}{@{}l l c c S[table-format=1.8(2)] S[table-format=1.7(2)]@{}}
    \toprule
    & & \multicolumn{2}{c}{$\sigma_\mathrm{fid}$ [pb]} & & \\
    \cmidrule(lr){3-4}
    {$\sqrt s$ [GeV]} & {final state} & {\ACRAIC} & {\ACRAIC strict} & {strict/\ACRAIC} & {strict/\YFSNLOEW} \\
    \midrule
    \multirow{7}{*}{91.19} & $\mu^+\mu^-$ & 1239.08(19) & 1238.18(19) & 0.9992734(4) & 1.003053(16) \\
     & $\tau^+\tau^-$ & 1246.92(22) & 1245.97(22) & 0.99923534(34) & 1.002698(18) \\
     & $e^+e^-$ & 1784.3(6) & 1783.7(6) & 0.9996582(18) & 1.00357(6) \\
     & $\nu\bar\nu$ & 2915.11(29) & 2910.33(29) & 0.99836233(19) & 1.0002246(9) \\
     & $d\bar d$ & 5513.0(14) & 5506.3(13) & 0.9987702(5) & 1.001099(14) \\
     & $u\bar u$\, & 4244.9(4) & 4241.4(4) & 0.99917(14) & 1.00002(14) \\
     & $\gamma\gamma$ & 27.6342(27) & 27.6019(27) & 0.99883(14) & 0.99836(14) \\
    \midrule
    & & \multicolumn{2}{c}{$\sigma_\mathrm{fid}$ [fb]} & & \\
    \cmidrule(lr){3-4}
    {$\sqrt s$ [GeV]} & {final state} & {\ACRAIC} & {\ACRAIC strict} & {strict/\ACRAIC} & {strict/\YFSNLOEW} \\
    \midrule
    \multirow{7}{*}{240} & $H\mu^+\mu^-$ & 4.732(9) & 4.732(9) & 1.000035(6) & 1.00196(25) \\
     & $H\nu\bar\nu$ & 12.912(6) & 12.879(6) & 0.9974617(16) & 0.999852(15) \\
     & $\mu^+\mu^-\tau^+\tau^-$ & 1.750(4) & 1.747(4) & 0.999(4) & 0.985(5) \\
     & $\tau^+\tau^-\tau^+\tau^-$ & 1.056(6) & 1.047(5) & 0.992(7) & 1.017(8) \\
     & $\tau^+\tau^-\nu_\mu\bar\nu_\mu$ & 3.610(13) & 3.602(13) & 0.998(5) & 1.001(4) \\
     & $\gamma\gamma$ & 3974.3(4) & 3968.6(4) & 0.99857(15) & 0.99793(15) \\
    \bottomrule
  \end{tabular}
  \caption{The strict $\mathcal O(\alpha)$ truncation of \ACRAIC, with the virtual on $\Coh_0$ only and no $(v/2)^2$ term, compared with the \ACRAIC $\mathcal O(\alpha)$ column of \TabRef{tab:fid}, which has the virtual on the whole $\mathcal O(\alpha)$ amplitude, and with \YFSNLOEW, all on the same events. The ratio errors are as in \TabRef{tab:fid}.}
  \label{tab:fid-strict}
\end{table*}

\begin{figure}
  \centering
  \includegraphics[width=0.48\textwidth]{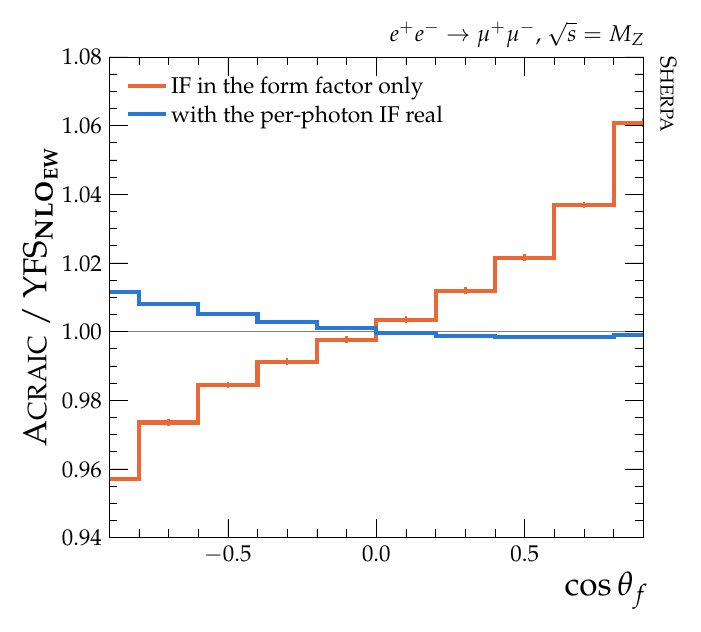}
  \caption{The effect of initial-final interference on the muon angle, with and without the rewighting.}
  \label{fig:ifi}
\end{figure}

\ACRAIC carries the interference between initial- and final-state emission in
its coherent partition sum, photon by photon. The naive \YFSNLOEW keeps it
only in the initial-final form factor, which is integrated over the photon
phase space and therefore over its direction. The
interference is antisymmetric in the fermion angle, so its integral survives
in the cross section but its angular dependence is lost. \YFSNLOEW can instead
reweight every photon by the ratio of the full eikonal factor, initial-final
terms included, to the one it was generated with. We label this variant
``+ IF real'' and it is the one used for \TabRef{tab:fid}. Against \ACRAIC on
the five $Z$-pole final states with charged fermions, the forward-backward
asymmetries agree to $0.004$ with it and differ by about $0.02$ without it, with
the sign of the charge product.
We show the effect on the muon angle in \FigRef{fig:ifi},
with and without the rewighting.

\label{sec:rad}

For processes with photons in the hard process the Born amplitude already
carries a hard photon, and the residuals and the partition sum are built on top
of it. \TabRef{tab:rad} compares the two columns for this family, which
comprises the $2\to3$ processes at the $Z$ pole, $\gamma\gamma\gamma$, and
$Hf\bar f\gamma$ at 240~\UGeV. The fiducial region requires at least one photon
with
\[
  \renewcommand{\arraystretch}{1.5}
  \begin{array}{@{}ll@{}}
    Z\text{ pole:}             & E_\gamma>0.2\sqrt s,\quad 60\,\UGeV<m_{ff}\ \text{(a visible pair)}, \\
    240~\UGeV\text{:}          & E_\gamma>10~\UGeV,\quad 66<m_{\mu\mu}<116~\UGeV\,,                   \\
    \text{both:}               & |\cos\theta_\gamma|<0.95,\quad \theta_{\gamma f}>0.12,               \\
    \gamma\gamma\gamma\text{:} & E_{\gamma_i}>0.3\sqrt s,\quad |\cos\theta_{\gamma_i}|<0.85\,
  \end{array}
\]
with the pair selected as in \TabRef{tab:fid}, and $\nu\bar\nu\gamma$, with no
visible pair, and $\gamma\gamma\gamma$, with no fermion pair, exempt from the
$m_{ff}$ floor.
The columns agree to $0.2$--$0.3\%$ for every channel of this
family (\TabRef{tab:rad}).

\begin{table*}
  \centering\small
  \sisetup{table-align-uncertainty=true}
  \begin{tabular}{@{}l l c c c S[table-format=1.4(2)]@{}}
    \toprule
    & & \multicolumn{3}{c}{$\sigma_\mathrm{fid}$ [pb]} & \\
    \cmidrule(lr){3-5}
    {$\sqrt s$ [GeV]} & {final state} & {\YFSLO} & {\YFSNLOEW} & {\ACRAIC} & {\ACRAIC/\YFSNLOEW} \\
    \midrule
    \multirow{7}{*}{91.19} & $\mu^+\mu^-\gamma$ & 5.8623(29) & 6.1907(33) & 6.2088(33) & 1.0029(8) \\
     & $\tau^+\tau^-\gamma$ & 5.5791(40) & 5.8401(45) & 5.8547(46) & 1.0025(11) \\
     & $e^+e^-\gamma$ & 7.6120(46) & 8.3337(56) & 8.3570(56) & 1.0028(9) \\
     & $\nu\bar\nu\gamma$ & 0.02827(8) & 0.029420(15) & 0.029439(15) & 1.0007(7) \\
     & $d\bar d\gamma$ & 3.1396(20) & 3.2884(22) & 3.2950(22) & 1.0020(9) \\
     & $u\bar u\gamma$ & 9.2010(48) & 9.5156(53) & 9.5175(53) & 1.0002(8) \\
     & $\gamma\gamma\gamma$ & 0.004270(4) & 0.004411(4) & 0.004413(4) & 1.0004(14) \\
    \midrule
    \multirow{2}{*}{240\,\textsuperscript{\dag}} & $H\mu^+\mu^-\gamma$ & 0.11049(15) & 0.11356(20) & 0.11291(17) & 0.9943(23) \\
     & $H\nu\bar\nu\gamma$ & 0.10279(7) & 0.10610(8) & 0.10624(8) & 1.0013(10) \\
    \bottomrule
  \end{tabular}
  \caption{Fiducial cross sections with a Born-level photon at $\mathcal O(\alpha)$, in the fiducial region of \SecRef{sec:rad}. \textsuperscript{\dag}\,Cross sections in fb. \YFSLO is the \YFS weight of the same events at $\bar\beta_0$ level.}
  \label{tab:rad}
\end{table*}

\subsection{The \texorpdfstring{$\mathcal O(\alpha^2)$}{O(alpha^2)} step and the perturbative uncertainty}\label{sec:nnlo}

The second-order predictions contain the photonic corrections of $\mathcal
  O(\alpha^2)$ but not the two-loop $\gamma Z$ boxes or the two-loop weak
corrections (\TabRef{tab:summary}), and we refer to them as photonic $\mathcal
  O(\alpha^2)$. We denote the $\mathcal O(\alpha^2)$ extension of \YFSNLOEW by
\YFSnNLO. With the real-virtual and the double real both columns, \ACRAIC and
\YFSnNLO, are carried to $\mathcal O(\alpha^2)$ on the same events, and the
step from $\mathcal O(\alpha)$ to $\mathcal O(\alpha^2)$ tests whether the two
matchings converge. \TabRef{tab:nnlo} gives the result for the $Z$-pole final
states and for Higgs production at 240~\UGeV. For $\gamma\gamma$ the fiducial
steps of the two columns agree, $+0.36\%$ for \ACRAIC and $+0.41\%$ for
\YFSnNLO, although \YFSnNLO has no double virtual at all for this
final state (\TabRef{tab:nnlo}, footnote b), so the two steps are not
contributions of quite the same physical content.

\begin{table*}
  \centering\small
  \begin{tabular}{@{}l l c c c c c c@{}}
    \toprule
    & & \multicolumn{2}{c}{\YFSNLOEW\ \ $\to$\ \ \YFSnNLO} &
        \multicolumn{2}{c}{\ACRAIC\ $\mathcal O(\alpha)\to\mathcal O(\alpha^2)$} &
        \multicolumn{2}{c}{\ACRAIC/\YFS} \\
    \cmidrule(lr){3-4}\cmidrule(lr){5-6}\cmidrule(lr){7-8}
    $\sqrt s$ [GeV] & final state & $\mathcal O(\alpha)$ & $\mathcal O(\alpha^2)$ & $\mathcal O(\alpha)$ & $\mathcal O(\alpha^2)$
      & $\mathcal O(\alpha)$ & $\mathcal O(\alpha^2)$ \\
    \midrule
    \multirow{6}{*}{91.19} & $\mu^+\mu^-$ [pb] & 1234.41(18) & 1236.48(18) & 1239.08(19) & 1241.01(19) & 1.003782(16) & 1.003660(16) \\
     & $\tau^+\tau^-$ [pb] & 1242.61(21) & 1244.67(21) & 1246.92(22) & 1248.59(22) & 1.003466(18) & 1.003150(17) \\
     & $e^+e^-$ [pb] & 1777.4(6) & 1781.9(7) & 1784.3(6) & 1789.5(8) & 1.00391(6) & 1.00424(20) \\
     & $\nu\bar\nu$ [pb] & 2909.68(29) & 2914.12(29) & 2915.11(29) & 2918.82(29) & 1.0018653(10) & 1.0016116(11) \\
     & $d\bar d$ [pb] & 5500.2(13) & 5508.6(25) & 5513.0(14) & 5518.4(14) & 1.002332(14) & 1.0018(4) \\
     & $\gamma\gamma$ [pb]\,\textsuperscript{b} & 27.639(7) & 27.754(7) & 27.626(7) & 27.708(7) & 0.9995318(28) & 0.998338(9) \\
    \midrule
    \multirow{2}{*}{240} & $H\mu^+\mu^-$ [fb]\,\textsuperscript{a} & 4.723(9) & 4.723(9) & 4.732(9) & 4.737(9) & 1.00192(25) & 1.00280(23) \\
     & $H\nu\bar\nu$ [fb]\, & 12.881(6) & 12.889(6) & 12.912(6) & 12.924(6) & 1.002397(15) & 1.002737(27) \\
    \bottomrule
  \end{tabular}
  \caption{The fiducial cross sections of \YFSNLOEW at $\mathcal O(\alpha)$ and \YFSnNLO at $\mathcal O(\alpha^2)$,
  including the \YFS double virtual for the $2\to2$ rows, and of \ACRAIC at $\mathcal O(\alpha)$, with the resonant exponentiation and the virtual on the whole $\mathcal O(\alpha)$ amplitude, and at $\mathcal O(\alpha^2)$, with the real-virtual and double reals and the photonic two-loop $\bet{0}^{(2)}$, and the ratio of the columns at each order. The $Z$-pole columns at $\mathcal O(\alpha^2)$ include the Dyson-resummed photon vacuum polarisation, the $Hf\bar f$ ones do not. \textsuperscript{a}\,$Hf\bar f$: \YFSnNLO without its double virtual. \textsuperscript{b}\,$\gamma\gamma$: there is no double virtual for this final state present in \YFSnNLO. It is present in \ACRAIC}.
  \label{tab:nnlo}
\end{table*}

For $Hf\bar f$ the \YFSnNLO column omits its own double virtual
(\TabRef{tab:nnlo}, footnote a), so the two ratios it enters, for $H\mu^+\mu^-$
and for $H\nu\bar\nu$, compare \ACRAIC's step against an incomplete \YFSnNLO
prediction and should be read with that caveat. With the virtual on the whole
$\mathcal O(\alpha)$ amplitude in both columns , each $\mathcal O(\alpha)$
column already contains the factorizable real-virtual, and the step contains
only the non-factorizable real-virtual, the double real and the two-loop
virtual, $\bet{0}^{(2)}$ in \ACRAIC and, for the $2\to2$ rows, the \YFS double
virtual in \YFSnNLO. At the $Z$ pole it also contains the Dyson-resummed photon
vacuum polarization.

\begin{figure*}
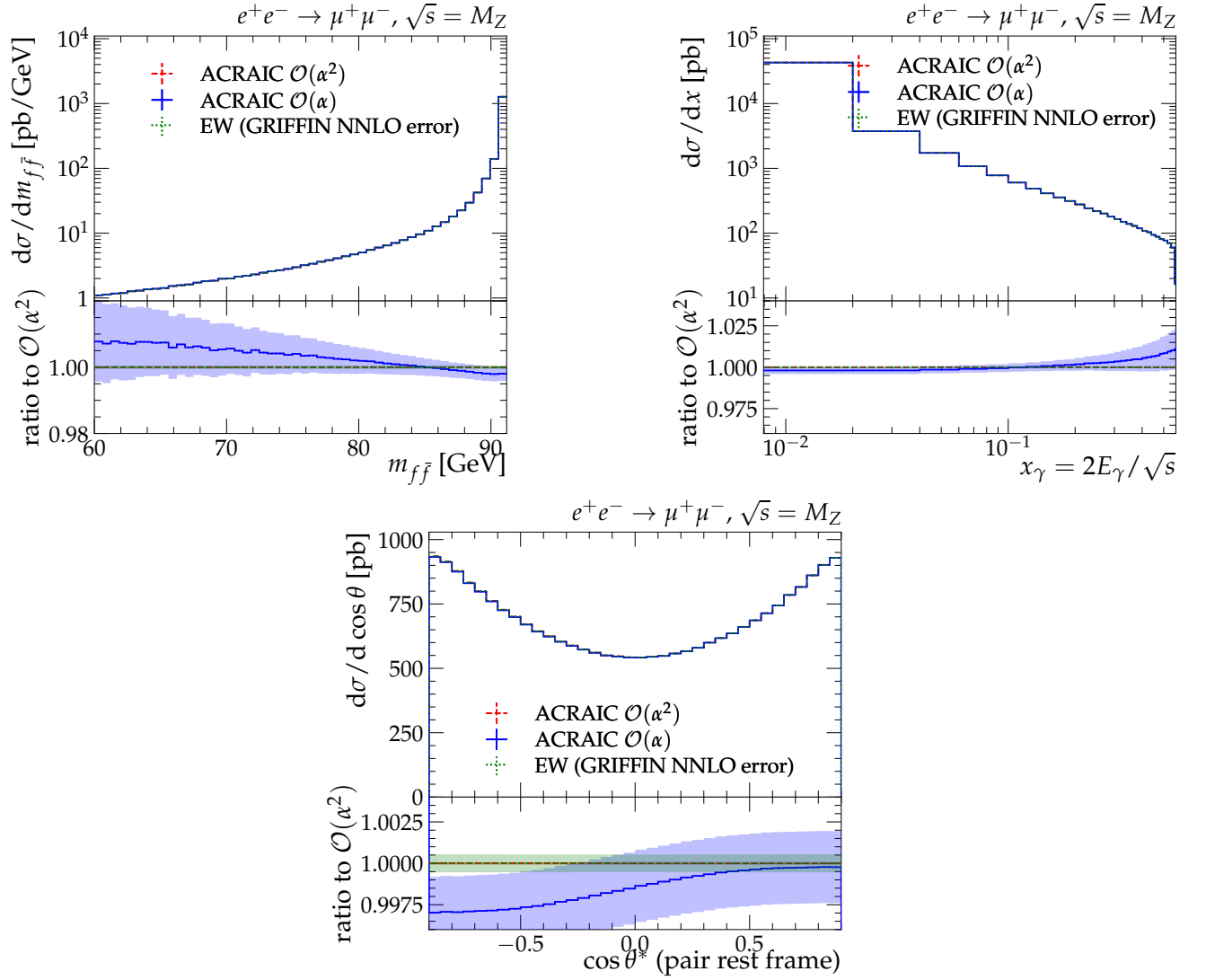

  \centering
  \newcommand{\tnpplot}[2]{\includegraphics[width=0.45\textwidth]{figures/validation/tnp/#1/CRAIC_EE/#2.pdf}}
  \tnpplot{zpole_mumu}{m_ff_win}\hfill
  \tnpplot{zpole_mumu}{x_gamma_hard}\hfill
  \tnpplot{zpole_mumu}{cos_star}\\[1ex]
  \caption{\ACRAIC at photonic $\mathcal O(\alpha^2)$ (red, with $\bet{0}^{(2)}$) and at
    $\mathcal O(\alpha)$ (blue), each with its theory
    nuisance-parameter band, and the electroweak band
    (green), in the
    fiducial region of \TabRef{tab:nnlo}, as a ratio to $\mathcal O(\alpha^2)$.
    We show the pair mass below the
    peak, the energy fraction $x_\gamma=2E_\gamma/\sqrt s$ of the hardest photon
    and the polar angle in the pair rest frame in $e^+e^-\to\mu^+\mu^-$.}
  \label{fig:tnp}
\end{figure*}

As an example application of the theory-nuisance-parameter band to $Z$-pole
precision distributions, \FigRef{fig:tnp} shows the differential distributions
of \ACRAIC at photonic $\mathcal O(\alpha^2)$ and at $\mathcal O(\alpha)$, each with the band of the theory nuisance parameters, and the
electroweak band.
The $\mathcal O(\alpha)$ band overlaps the $\mathcal O(\alpha^2)$ prediction in
every bin of the pair mass below the peak for $\mu^+\mu^-$,
of the hardest-photon energy for $\mu^+\mu^-$. In the backward $\mu^-$ angle the step slightly
exceeds it.
The two-loop photonic $\bet{0}^{(2)}$ is in the \ACRAIC $\mathcal O(\alpha^2)$
column of \TabRef{tab:nnlo}. \TabRef{tab:b02} gives its pieces for $\mu^+\mu^-$
at the $Z$ pole, each measured on the same events with and without it, and
compares the step with \kkmc's. The leading- and next-to-leading-logarithmic
step of the photonic vertex form factors is the one \kkmc adds between
\OCEEX{1} and \OCEEX{2}, and the two agree. The product of the initial- and
final-state one-loop vertices is contained in \kkmc's \OCEEX{1} through
$(1+F_i)(1+F_f)$ but not in our $\mathcal O(\alpha)$ columns, so it enters our
step only. The exact photonic two-loop terms, the constant of the two-loop
vertex~\cite{Penin:2005eh} beyond its logarithms and the non-factorizable
two-loop boxes of photon exchange from the helicity amplitudes
of~\cite{Dave:2024ewl}, are not in \kkmc. For $\gamma\gamma$ the complete
two-loop hard function \EqRef{eq:diphoton-2l} changes the cross section by
$-3\times10^{-5}$. They are an order of magnitude below the other
pieces, nearly uniform in angle and in $m_{ff}$, and the same in both columns.
With the $\mathcal O(\alpha)$ column defined as \kkmc defines it, with the
virtual on $\Coh_0$, the two steps agree with and without $\bet{0}^{(2)}$. With
the virtual on the whole $\mathcal O(\alpha)$ amplitude they differ by the
factorizable virtual on $\bet{1}$, which \ACRAIC evaluates at the Born point
and \kkmc at the invariants of the partition.

\begin{table}
  \centering\small
  \begin{tabular}{@{}lcc@{}}
    \toprule
                                                 & \ACRAIC   & \kkmc                             \\
    \midrule
    \multicolumn{3}{@{}l}{$\bet{0}^{(2)}$: shift of $\sigma_{\rm fid}$ [$10^{-3}$]}              \\
    photonic vertices, LL and NLL                & $+1.160$  & $+1.154$                          \\
    initial-state $\times$ final-state vertices  & $+0.764$  & in \OCEEX{1}                      \\
    initial--final pairs, photonic $\times$ weak & $+0.103$  & --                                \\
    exact two-loop vertex constant and boxes     & $-0.062$  & --                                \\
    \midrule
    \multicolumn{3}{@{}l}{$\mathcal O(\alpha)\to\mathcal O(\alpha^2)$ step, virtual on $\Coh_0$} \\
    with $\bet{0}^{(2)}$ (LL and NLL)            & $0.99964$ & $0.99961$                         \\
    without $\bet{0}^{(2)}$                      & $0.99848$ & $0.99846$                         \\
    \multicolumn{3}{@{}l}{$\mathcal O(\alpha)\to\mathcal O(\alpha^2)$ step, virtual on $\Coh_1$} \\
    with $\bet{0}^{(2)}$                         & $1.0016$  & $1.0013$                          \\
    \bottomrule
  \end{tabular}
  \caption{The two-loop photonic $\bet{0}^{(2)}$ for $e^+e^-\to\mu^+\mu^-$ at
    the $Z$ pole in the fiducial region of \TabRef{tab:nnlo}. Top: its pieces,
    each the shift of the \ACRAIC fiducial cross section on the same events
    with and without it, against the step \kkmc adds between \OCEEX{1} and \OCEEX{2}.
    Bottom: the $\mathcal O(\alpha)\to\mathcal O(\alpha^2)$ step of \ACRAIC and
    of \kkmc with the $\mathcal O(\alpha)$ virtual on the Born-level amplitude
    only, as in \kkmc's \OCEEX{1}, and on the whole $\mathcal O(\alpha)$
    amplitude.}
  \label{tab:b02}
\end{table}

For $\nu\bar\nu$, with initial-state radiation only, both columns move by
$1.3$--$1.5\times10^{-3}$, most of it from the two-loop virtual, and the
$0.2\%$ difference of $\mathcal O(\alpha)$ remains, $0.16\%$ at $\mathcal
  O(\alpha^2)$.

\TabRef{tab:unc} gives the budget of the fiducial cross sections at the $Z$
pole. At $\mathcal O(\alpha)$ the band is $1.5$--$2.2\times10^{-3}$ for
$\mu^+\mu^-$ and $\tau^+\tau^-$, somewhat lower, $1.5\times10^{-3}$ or below,
for $\nu\bar\nu$ and $d\bar d$, growing with the final-state logarithm through the
cross terms of $\bet{0}$ ($T_{\bet{0}}\simeq2$), and nearly $4\times10^{-3}$ for
$e^+e^-$, whose $t$ channel adds a final-state $\bet{0}$ and $\bet{1}$ of the
initial-state size. The complete step, $1.0$--$1.6\times10^{-3}$
for $\mu^+\mu^-$ and $\tau^+\tau^-$, about $1.3\times10^{-3}$ or below for
$\nu\bar\nu$ and $d\bar d$, and $2.9\times10^{-3}$ for $e^+e^-$,
lies at $|\theta|\approx0.65$--$0.9$. Most of it is $\bet{0}^{(2)}$, at
$|\theta|\approx0.8$--$0.9$ against the $\bet{0}$ band for $\mu^+\mu^-$,
$\tau^+\tau^-$ and $\nu\bar\nu$, $0.65$ for $d\bar d$ and $1.05$ for $e^+e^-$.
The rest is at $|\theta|\lesssim1$ against the other classes, except for
$\nu\bar\nu$, where it is $1.4$. For
charged final states the other classes are the final-state $\bet{1}$. The
initial-state $\bet{1}$ template integrates to a few $10^{-6}$ over the fiducial
region, even with an independent parameter in each bin of $x_\gamma$, so the
cancellation of a template between regions of phase space does not matter here.
With $\bet{0}^{(2)}$ in the prediction the $\mathcal O(\alpha^2)$ band is a few
$10^{-5}$, and $1.6\times10^{-4}$ for $e^+e^-$, from $\bet{0}^{(3)}$ and for
$\mu^+\mu^-$ mostly its cross terms. The $\bet{1}$ and $\bet{2}$ classes
contribute below $10^{-5}$, except $\bet{1}$ for $e^+e^-$ at $4\times10^{-5}$.

\begin{table*}
  \centering\small
  \begin{tabular}{@{}l c c c c c c @{}}
    \toprule
    $\sqrt s=M_Z$                                  & $\mu^+\mu^-$   & $\tau^+\tau^-$ & $\nu\bar\nu$ & $e^+e^-$   & $u\bar u$\,\textsuperscript{*} & $d\bar d$  \\
    \midrule
    $\sigma_{\rm fid}$ [pb]                        & 1241.0(2)      & 1248.6(2)      & 2918.8(3)    & 1789.5(8)  & 4244.9(4)                      & 5518.4(14) \\
    \multicolumn{7}{@{}l}{$\mathcal O(\alpha)$ band [$10^{-3}$]}                                                                                               \\
    \quad $\bet{0}$ initial                        & 1.43           & 1.44           & 1.45         & 1.42       & 1.44                           & 1.45       \\
    \quad $\bet{0}$ final                          & 0.42           & 0.13           & 0            & 1.42       & 0.41                           & 0.089      \\
    \quad $\bet{0}$ initial$\times$final           & 1.55           & 0.85           & 0            & 2.84       & 1.03                           & 0.24       \\
    \quad $\bet{1}$ initial                        & 0.00492        & 0.00603        & 0.11         & 0.27       & 0.0263                         & 0.0728     \\
    \quad $\bet{1}$ final                          & 0.38           & 0.0964         & 0            & 1.38       & 0.38                           & 0.0736     \\
    \quad $\bet{2}$ missing                        & 0.03           & 0.0119         & $<0.001$     & 0.16       & --                             & 0.00148    \\
    \quad total                                    & 2.18           & 1.68           & 1.46         & 3.75       & 1.85                           & 1.48       \\
    step to $\mathcal O(\alpha^2)$ [$10^{-3}$]     & 1.56           & 1.34           & 1.27         & 2.91       & --                             & 0.97       \\
    \quad $|\theta|$                               & 0.71           & 0.79           & 0.88         & 0.77       & --                             & 0.66       \\
    \quad of it $\bet{0}^{(2)}$ [$10^{-3}$]        & $+1.94(1)$     & $+1.40(1)$     & $+1.12(1)$   & $+3.67(1)$ & --                             & $+0.96(1)$ \\
    \midrule
    \multicolumn{7}{@{}l}{$\mathcal O(\alpha^2)$ band, with $\bet{0}^{(2)}$ [$10^{-3}$]}                                                                       \\
    \quad $\bet{0}$                                & 0.0691         & 0.0394         & 0.026        & 0.16       & --                             & 0.0269     \\
    \quad $\bet{1}$                                & 0.00557        & $<0.001$       & 0.00283      & 0.0377     & --                             & 0.00245    \\
    \quad $\bet{2}$                                & $<0.001$       & $<0.001$       & $<0.001$     & 0.00417    & --                             & $<0.001$   \\
    \quad total                                    & 0.0693         & 0.0395         & 0.0262       & 0.16       & --                             & 0.027      \\
    EW higher orders [$10^{-3}$]                   & 2.09           & --             & --           & --         & --                             & --         \\
    EW input, $\alpha(M_Z)$ vs $G_\mu$ [$10^{-3}$] & $-9.4(3)$      & $-9.5(4)$      & $-9.2(2)$    & $-8.8(9)$  & $-7.6(2)$                      & $-10.0(6)$ \\
    \quad at $\mathcal O(\alpha)$ [$10^{-3}$]      & $-7.0(3)$      & $-7.5(4)$      & $-7.6(2)$    & $-4.6(7)$  & --                             & $-8.2(5)$  \\
    initial-state pairs [$10^{-3}$]                & $-2.54\pm0.22$ & --             & --           & --         & --                             & --         \\
    \bottomrule
  \end{tabular}
  \caption{Uncertainty budget of the fiducial cross sections at the $Z$ pole,
  in units of $10^{-3}$ of $\sigma_{\rm fid}$: the size of each theory
  nuisance parameter, at $\mathcal O(\alpha)$
  and at $\mathcal O(\alpha^2)$ with the photonic $\bet{0}^{(2)}$ in
  the prediction, the complete $\mathcal O(\alpha)\to\mathcal O(\alpha^2)$
  step on the same events with its $|\theta|$ against the $\mathcal O(\alpha)$
  band and the part of it from $\bet{0}^{(2)}$, the \Griffin two-loop
  electroweak step on the same events, $\sigma_{\rm fid}$ in the $\alpha(M_Z)$
  scheme over the $G_\mu$ minus one
  and the initial-state light-fermion pair correction of~\cite{Arbuzov:1999xso},
   which \ACRAIC does not generate. The multi-photon remainder vanishes for these $2\to2$ processes.}
  \label{tab:unc}
\end{table*}

\begin{figure}
  \centering
  \includegraphics[width=0.48\textwidth]{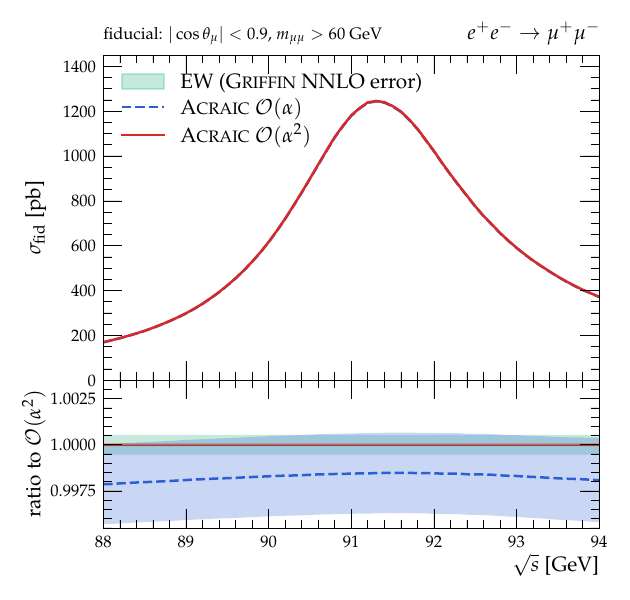}
  \caption{The $Z$ line shape of $e^+e^-\to\mu^+\mu^-$ in the fiducial region
    of \TabRef{tab:nnlo}: \ACRAIC at photonic $\mathcal O(\alpha^2)$ (red) and
    $\mathcal O(\alpha)$ (blue) with their theory-nuisance-parameter bands, both
    with the \Griffin two-loop electroweak step applied, and the electroweak
    band, the estimated missing higher orders of that correction
    (green), on
    $\mathcal O(\alpha^2)$, at energies $0.1$~\UGeV apart. Below, the ratio to
    $\mathcal O(\alpha^2)$. All curves at one energy come from the same events.}
  \label{fig:zscan}
\end{figure}

\FigRef{fig:zscan} repeats this across the $Z$ line shape,
$88\le\sqrt s\le94~\UGeV$, for $e^+e^-\to\mu^+\mu^-$ in the fiducial region of
\TabRef{tab:nnlo}, with $\bet{0}^{(2)}$ in the $\mathcal O(\alpha^2)$
prediction. The $\mathcal O(\alpha)$ band is about $2.2\times10^{-3}$ across the
line shape and the complete step $1.5$--$2.1\times10^{-3}$, at $|\theta|$ from
$0.7$ at the peak to $1$ at $88~\UGeV$, where $\bet{0}^{(2)}$ grows to the size
of the $\bet{0}$ band. The $\mathcal O(\alpha^2)$ band is about $7\times10^{-5}$.
At $240~\UGeV$ (\TabRef{tab:unc240}) the $\mathcal O(\alpha)$
band is $1.7$--$2.7\times10^{-3}$ and the complete step about $10^{-3}$, at
$|\theta|\approx0.35$ for $H\mu^+\mu^-$ and $0.6$ for $H\nu\bar\nu$. The
step is almost entirely $\bet{0}^{(2)}$ for $H\mu^+\mu^-$ and half of it for
$H\nu\bar\nu$. The $\mathcal O(\alpha^2)$ band is $10^{-4}$ or below. The
electroweak band of $H\mu^+\mu^-$ (\SecRef{sec:uncertainty}), half the scheme
spread of $-1.6(4)\times10^{-2}$ in \TabRef{tab:unc240}, is about
$8\times10^{-3}$, larger than the $\mathcal O(\alpha^2)$ QED band and of the size of the known
next-to-next-to-leading-order corrections to $e^+e^-\to ZH$.

\begin{table}[!htb]
  \centering\small
  \begin{tabular}{@{}l c c @{}}
    \toprule
    $\sqrt s=240$ GeV                              & $H\mu^+\mu^-$ & $H\nu\bar\nu$       \\
    \midrule
    $\sigma_{\rm fid}$ [fb]                        & 4.737(9)      & 12.924(6)           \\
    \multicolumn{3}{@{}l}{$\mathcal O(\alpha)$ band [$10^{-3}$]}                         \\
    \quad $\bet{0}$ initial                        & 1.70          & 1.70                \\
    \quad $\bet{0}$ final                          & 0.56          & 0                   \\
    \quad $\bet{0}$ initial$\times$final           & 1.95          & 0                   \\
    \quad $\bet{1}$ initial                        & 0.32          & 0.0303              \\
    \quad $\bet{1}$ final                          & 0.30          & 0                   \\
    \quad $\bet{2}$ missing                        & 0.0681        & 0.32                \\
    \quad total                                    & 2.68          & 1.73                \\
    step to $\mathcal O(\alpha^2)$ [$10^{-3}$]     & 0.94          & 0.97                \\
    \quad $|\theta|$                               & 0.35          & 0.56                \\
    \quad of it $\bet{0}^{(2)}$ [$10^{-3}$]        & $+1.01(1)$    & $+0.55(1)$          \\
    \midrule
    \multicolumn{3}{@{}l}{$\mathcal O(\alpha^2)$ band, with $\bet{0}^{(2)}$ [$10^{-3}$]} \\
    \quad $\bet{0}$                                & 0.0957        & 0.033               \\
    \quad $\bet{1}$                                & 0.0105        & $<0.001$            \\
    \quad $\bet{2}$                                & 0.00198       & 0.0092              \\
    \quad total                                    & 0.0963        & 0.0343              \\
    EW higher orders [$10^{-3}$]                   & --            & --                  \\
    EW input, $\alpha(M_Z)$ vs $G_\mu$ [$10^{-3}$] & $-16(4)$      & $-16(1)$            \\
    \quad at $\mathcal O(\alpha)$ [$10^{-3}$]      & $-12(4)$      & $-14(1)$            \\
    initial-state pairs [$10^{-3}$]                & --            & --                  \\
    \bottomrule
  \end{tabular}
  \caption{Uncertainty budget of the fiducial cross sections at 240~\UGeV,
  in units of $10^{-3}$ of $\sigma_{\rm fid}$: the size of each theory
  nuisance parameter, at $\mathcal O(\alpha)$
  and at $\mathcal O(\alpha^2)$ with the photonic $\bet{0}^{(2)}$ in the
  prediction, the complete $\mathcal O(\alpha)\to\mathcal O(\alpha^2)$ step
  on the same events with its $|\theta|$ against the $\mathcal O(\alpha)$
  band and the part of it from $\bet{0}^{(2)}$, and the electroweak-input ambiguity, $\sigma_{\rm fid}$
  in the $\alpha(M_Z)$ scheme over the $G_\mu$.}
  \label{tab:unc240}
\end{table}

\begin{figure*}
  \centering
  \includegraphics[width=0.48\textwidth]{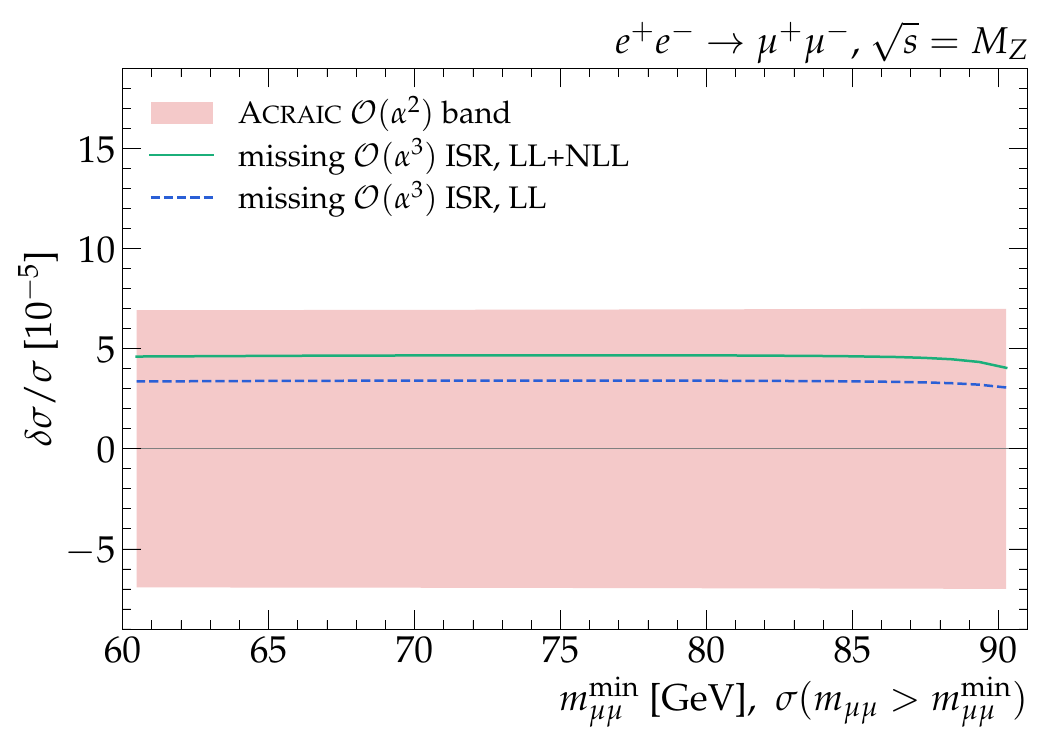}\hfill
  \includegraphics[width=0.48\textwidth]{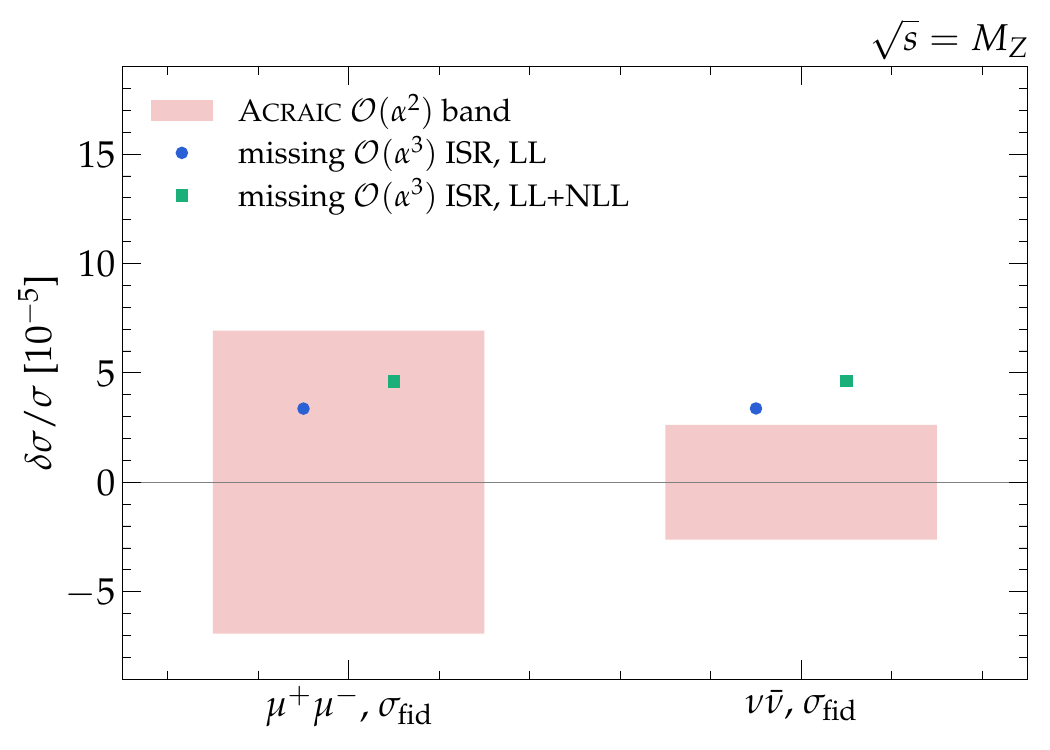}
  \caption{The \ACRAIC $\mathcal O(\alpha^2)$ theory-nuisance-parameter band
  (red) and the initial-state terms of $\mathcal O(\alpha^3)$ that an
  exponentiated $\mathcal O(\alpha^2)$ prediction misses, at leading (LL) and
  next-to-leading (NLL) logarithmic accuracy, from the radiators
  of~\cite{Blumlein:2020jrf,Ablinger:2020qvo}, relative to the cross section
  at $\sqrt s=M_Z$. Left, $e^+e^-\to\mu^+\mu^-$ with
  $m_{\mu\mu}>m_{\mu\mu}^{\min}$, in the fiducial region of \TabRef{tab:nnlo}
  otherwise. Right,
  the fiducial cross-sections for $\mu^+\mu^-$ and $\nu_\mu\bar\nu_\mu$.}
  \label{fig:tnp-isr}
\end{figure*}

The initial-state part of the terms the $\mathcal O(\alpha^2)$ band stands for
can be computed. The complete photonic initial-state radiator of $\mathcal
  O(\alpha^2)$ is known, including the electron mass~\cite{Blumlein:2020jrf}. For
photon emission alone the renormalization-group structure of the
radiator~\cite{Ablinger:2020qvo} then fixes its leading ($\alpha^3L^3$) and
next-to-leading ($\alpha^3L^2$) logarithms at $\mathcal O(\alpha^3)$ from the
radiators of $\mathcal O(\alpha)$ and $\mathcal O(\alpha^2)$, with
$L=\ln(s/m_e^2)$. We subtract the part of these terms that the \YFS
exponentiation of an $\mathcal O(\alpha^2)$ prediction already contains, and
what remains is what a prediction like ours misses (\FigRef{fig:tnp-isr}). The
test covers initial-state radiation only. For $e^+e^-\to\mu^+\mu^-$ at the $Z$
pole the missing initial-state terms are $3.4\times10^{-5}$ at leading and
$4.6\times10^{-5}$ at next-to-leading logarithmic accuracy, nearly independent
of the lower cut on $m_{\mu\mu}$. The band of $6.9\times10^{-5}$ on
$\sigma_{\rm fid}$ contains them at $|\theta|\approx0.5$--$0.7$, but most of
that band is the initial$\times$final cross term of $\bet{0}$, which this test
does not probe. For $e^+e^-\to\nu\bar\nu$ there is only initial-state radiation
and the test is complete. There the band, $2.6\times10^{-5}$ on the total cross
section, is the soft leading-logarithmic part of $\bet{0}^{(3)}$, and the
missing terms, $3.4\times10^{-5}$ and $4.6\times10^{-5}$, are of the same size,
at $|\theta|\approx1.3$ and $1.8$. The band is set by the soft logarithms
because the initial-state $\bet{1}$ template integrates to almost zero in an
inclusive cross section, and the hard-photon and next-to-leading logarithms
enter through this test instead. The two estimates are complementary, and both
lie far below the pair and electroweak uncertainties of
\SecRef{sec:uncertainty}. In the radiative return to the $Z$ at higher energies
the missing $\mathcal O(\alpha^3)$ terms depend on how the exponentiation
distributes soft and hard photons, so the semi-analytical result is not a
unique reference there.

At the $Z$ pole the \Griffin two-loop electroweak step of
\SecRef{sec:uncertainty} is $1$--$2\times10^{-3}$ in $\sigma_{\rm fid}$ across
the scan, $2.1\times10^{-3}$ for $\mu^+\mu^-$ in \TabRef{tab:unc}, and about
$3\times10^{-3}$ in $A_{\rm FB}$ at $M_Z$, comparable to the $\mathcal
  O(\alpha)$ QED band and thirty times the $\mathcal O(\alpha^2)$ one.
\FigRef{fig:zscan} applies it as a correction, \TabRef{tab:unc} does not. With
the correction applied, the electroweak band of \FigRef{fig:zscan} is the
estimate of its missing higher orders, which is below the $\mathcal O(\alpha)$
photonic band at every energy. The input-scheme difference of \TabRef{tab:unc},
$\sigma_{\rm fid}$ in the $\alpha(M_Z)$ scheme against the $G_\mu$ one, is
larger, $-7.6$ to $-10\times10^{-3}$. The initial-state pair correction at the
peak, about $-2.5\times10^{-3}$, is larger than either the $\mathcal O(\alpha)$
photonic band or the \Griffin step. With the photonic corrections at $\mathcal
  O(\alpha^2)$, including $\bet{0}^{(2)}$, and the pair corrections
of~\cite{Arbuzov:1999xso}, the perturbative uncertainty of the $Z$ line shape
is electroweak, followed by that of the pair corrections.

\subsection{The Higgs recoil mass at FCC-ee}\label{sec:recoil}

As an example application of the theory-nuisance-parameter machinery of
\SecRef{sec:uncertainty}, rather than a further validation, we consider the
Higgs recoil mass, which a dedicated phenomenological study could develop into
a full precision analysis. At FCC-ee the Higgs boson mass will be measured from
the mass recoiling against the lepton pair in $e^+e^-\to Z(\to\ell^+\ell^-)H$,
independently of the Higgs decay. A precision comparable to the Higgs width,
about $4~\UMeV$, is the target, both for the electroweak fits and to probe the
electron Yukawa coupling in resonant Higgs
production~\cite{Azzurri:2021nmy,Eysermans:2026abl}. The recoil mass is
computed from the nominal collision energy, so every photon that escapes,
initial-state radiation above all, and the spread of the beam energies, move
events out of the peak at $M_H$. The line shape of the peak is a QED
prediction, and its theory uncertainty enters the mass measurement directly.

\begin{figure*}
  \centering
  \includegraphics[width=0.45\textwidth]{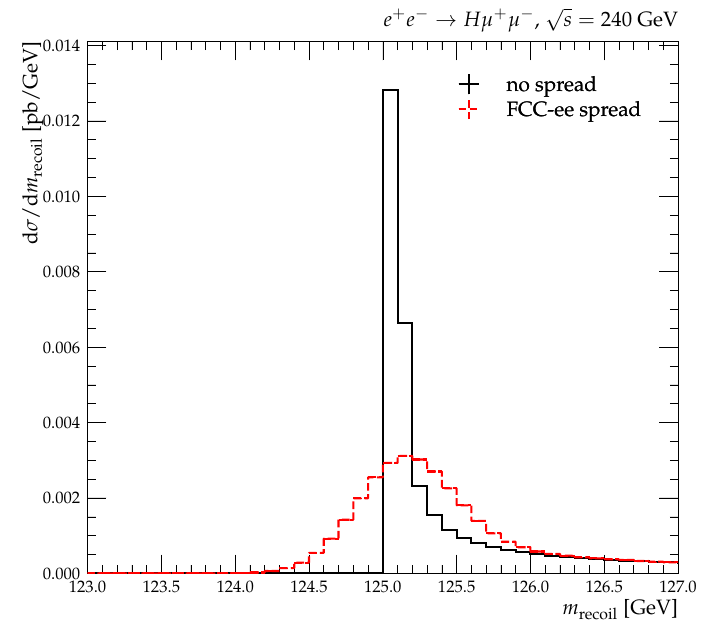}\hfill
  \includegraphics[width=0.45\textwidth]{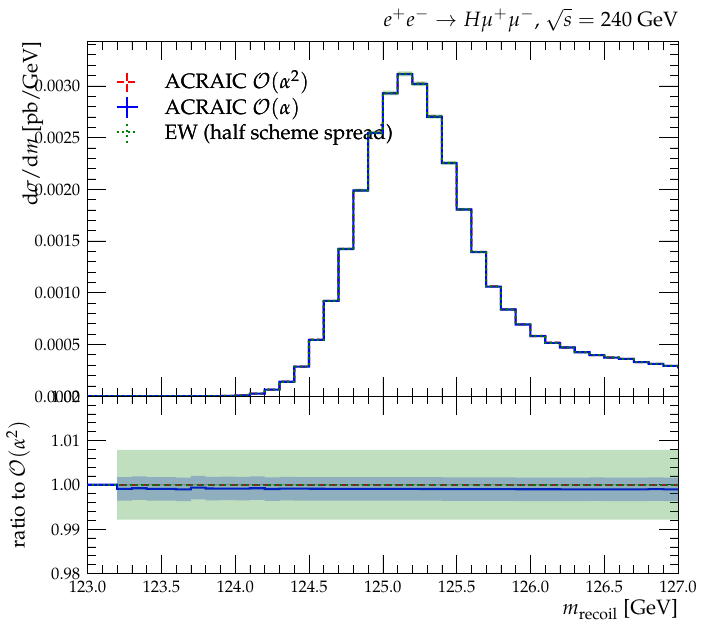}
  \caption{The recoil mass in $e^+e^-\to H\mu^+\mu^-$ at $\sqrt s=240~\UGeV$,
    \ACRAIC at photonic $\mathcal O(\alpha^2)$, in the fiducial region of
    \TabRef{tab:nnlo}. Left: without and with the beam-energy spread of FCC-ee
    ($0.165\%$ per beam\newinline{~\cite{FCC:2018evy}}). Right: with the spread, $\mathcal O(\alpha^2)$ and
    $\mathcal O(\alpha)$ on the same events with their theory-nuisance-parameter
    bands, and the electroweak band (green, half the QED-stripped scheme
    spread), as a ratio to $\mathcal O(\alpha^2)$.}
  \label{fig:recoil}
\end{figure*}

\ACRAIC describes it event by event. Every photon is generated exclusively, and
initial- and final-state emission and their interference are summed coherently
at amplitude level, with the soft logarithms exponentiated to all orders. For
$e^+e^-\to H\mu^+\mu^-$ the $\mathcal O(\alpha^2)$ prediction contains
\begin{itemize}
  \item the complete one-loop electroweak corrections to the Born
        (\SecRef{sec:virtual}) applied to the whole $\mathcal O(\alpha)$ amplitude,
  \item the exact tree amplitudes with one and with two photons, the latter through
        $\bet{2}$ of \EqRef{eq:beta2},
  \item the real-virtual, the one-loop correction to the emission of one photon,
        averaged over helicities (\SecRef{sec:oalpha2}),
  \item the photonic two-loop $\bet{0}^{(2)}$, described below,
  \item the resonant initial-final logarithm of the $Z$, exponentiated into its
        propagator (\EqRef{eq:resonant-log}),
  \item the beam-energy spread of FCC-ee (\AppRef{app:bes}).
\end{itemize}
The photonic $\bet{0}^{(2)}$ is a factor $\Virtual^{(2)}$ on the Born amplitude,
built from the infrared-finite photonic one-loop factors $f_{ij}$ of every pair
of charged legs, here the $e^+e^-$ pair, the $\mu^+\mu^-$ pair and the four
initial-final pairs. With $w$ the rest of the one-loop factor, the weak loops
and the vacuum polarization,
\begin{align}
  \Virtual_f     & =\sum_{i<j}f_{ij} ,
  \label{eq:delta-f}                                                                    \\
  \Virtual^{(2)} & =\tfrac12\,\Virtual_f^2+\Virtual_f\,w+\Virtual_{\rm NLL}+C_{\rm V} .
  \label{eq:beta02-content}
\end{align}
The first term carries the leading logarithms of the two-loop photonic
corrections, including the product of the initial- and final-state vertices.
The second is the product of the photonic and the weak one-loop corrections.
$\Virtual_{\rm NLL}$ is the next-to-leading logarithm, the two-loop collinear
anomalous dimension of each leg, and together with the first term it
reproduces the leading- and next-to-leading-logarithmic step validated in
\TabRef{tab:b02}.
$C_{\rm V}$ is the non-logarithmic constant of the massive two-loop vertex for
the $e^+e^-$ and the $\mu^+\mu^-$ pair~\cite{Penin:2005eh}.
The prediction does not contain
\begin{itemize}
  \item the non-factorizable two-loop boxes of photon exchange between the initial and
        the final state, which \ACRAIC includes for $2\to2$ processes only. For
        $\mu^+\mu^-$ at the $Z$ pole they and the vertex constant together shift
        $\sigma_{\rm fid}$ by less than $10^{-4}$ (\TabRef{tab:b02}),
  \item the remaining photonic two-loop terms without large logarithms, beyond the
        vertex constants, and the photonic corrections of third order, which the
        $\mathcal O(\alpha^2)$ band of $10^{-4}$ or below estimates
        (\SecRef{sec:uncertainty}),
  \item the two-loop electroweak and the mixed electroweak--QCD corrections, which the
        electroweak band of about $1\%$ estimates (\SecRef{sec:uncertainty}),
  \item light-fermion pairs.
\end{itemize}
\FigRef{fig:recoil} shows the peak for
$e^+e^-\to H\mu^+\mu^-$ at 240~\UGeV, the energy of the FCC-ee $ZH$ run. Without the spread the recoil mass has a
sharp edge at $M_H$ and the radiative tail above it. The FCC-ee spread
at the $ZH$ energy~\cite{FCC:2018evy}, $0.165\%$ per beam, smears the edge by about $0.3~\UGeV$ and moves a fifth of
the events below $M_H$, at an unchanged cross section. At the peak the photonic
higher orders are small. Between
$\mathcal O(\alpha)$ and $\mathcal O(\alpha^2)$, on the same events, the mean
recoil mass moves by a few $10^{-2}~\UMeV$, and the
theory-nuisance-parameter bands correspond to a few $10^{-2}~\UMeV$ at
$\mathcal O(\alpha)$ and $10^{-3}~\UMeV$ at $\mathcal O(\alpha^2)$, with and
without the spread. The missing weak corrections, about $1\%$
(\SecRef{sec:uncertainty}), change the normalization and not the position of
the peak. These estimates cover the matched photonic orders on top of the YFS
exponent common to both orders. They do not include light-fermion pairs,
beamstrahlung, an uncertainty of the beam-energy spread itself or a shape of the
missing weak corrections, and they indicate the size of the QED effects on the
peak position rather than a complete uncertainty of the mass. They are two
orders of magnitude below the $4~\UMeV$ target. In the recoil spectrum the
photonic uncertainty is largest in the hard-photon tail, where the second hard
initial-state photon enters (\FigRef{fig:tnp}).

\subsection{Four-lepton production at the \texorpdfstring{$W$}{W}-pair threshold}\label{sec:4f}

As an application beyond $2\to2$ and $2\to3$ processes we compute $e^+e^-\to4$
leptons at $\sqrt s=161~\UGeV$, the $W$-pair threshold, with the complete tree
amplitudes of each final state, resonant and non-resonant diagrams alike. The
$W$ bosons are stages of their own, so photons are emitted in the production
and in each decay, and the $W$ propagators are shifted per partition. The
one-loop virtual is the complete electroweak correction of the $2\to4$ process
from \OpenLoops, evaluated with massless leptons and massified as in
\EqRef{eq:massification}, and the real-virtual is factorized. The resonant
exponentiation of \EqRef{eq:resonant-log} is not applied, since it is not
derived for charged resonances. \ACRAIC is at $\mathcal O(\alpha)$, and the
electroweak inputs are those of \TabRef{tab:ew-params}. The fiducial region
follows the tuned comparison of the LEP2 four-fermion
report~\cite{Grunewald:2000ju}. Photons within $5^\circ$ of the beams are
invisible, and every other photon is recombined with the closest charged lepton
if their invariant mass is below $5~\UGeV$ or its energy below $1~\UGeV$.
Charged leptons must lie at least $20^\circ$ from the beams, opposite-sign
same-flavour lepton pairs must have a mass above $50~\UGeV$, and the missing
transverse momentum must exceed $25~\UGeV$.

\begin{table}[!htb]
  \centering\small
  \begin{tabular}{@{}l c c c@{}}
    \toprule
    final state & \YFSLO [fb] & \ACRAIC [fb] & \ACRAIC/\YFSLO \\
    \midrule
    $e^-\bar\nu_e\,\mu^+\nu_\mu$       & 19.30 & 20.19(6)  & 1.046 \\
    $\mu^-\bar\nu_\mu\,e^+\nu_e$       & 19.41 & 20.23(7)  & 1.042 \\
    $e^-\bar\nu_e\,\tau^+\nu_\tau$     & 19.29 & 20.35(6)  & 1.055 \\
    $\tau^-\bar\nu_\tau\,e^+\nu_e$     & 19.24 & 20.31(7)  & 1.056 \\
    $\mu^-\bar\nu_\mu\,\tau^+\nu_\tau$ & 19.85 & 21.41(17) & 1.079 \\
    $\tau^-\bar\nu_\tau\,\mu^+\nu_\mu$ & 19.84 & 21.58(8)  & 1.088 \\
    $\mu^+\mu^-\nu_\mu\bar\nu_\mu$     & 13.84 & 14.92(4)  & 1.079 \\
    \bottomrule
  \end{tabular}
  \caption{Fiducial cross sections of $e^+e^-\to4$ leptons at
    $\sqrt s=161~\UGeV$, \ACRAIC at $\mathcal O(\alpha)$ and the Born level of the
    same resummation, in the fiducial region of the text, with the statistical
    error of \ACRAIC.}
  \label{tab:4f}
\end{table}

\TabRef{tab:4f} gives the fiducial cross sections of the six charged-current
channels $\ell^-\bar\nu_\ell\,\ell'^+\nu_{\ell'}$ with $\ell\neq\ell'$ and of
$\mu^+\mu^-\nu_\mu\bar\nu_\mu$. The $\mathcal O(\alpha)$ correction relative to
the Born level of the resummation is $4$--$6\%$ for the channels with an
electron and $8$--$9\%$ for the others. The charge-conjugate channels, such as
$e^-\bar\nu_e\,\mu^+\nu_\mu$ and $\mu^-\bar\nu_\mu\,e^+\nu_e$, are generated
and evaluated independently, and they agree within their statistical errors.

These final states are also where the squared-level matching fails, and so they
motivate the amplitude-level construction. In the same samples the weights of
\YFSNLOEW have large single-event tails, and its fiducial cross sections carry
statistical errors from $2\%$ to nearly $100\%$ across the seven channels, so
they are set by a handful of events and we do not compare with it. We attribute
the tails to the soft final-state photons of the four-lepton system. The factor
$1+\Virtual_j$ of \YFSNLOEW divides the exact one-photon remainder by the
eikonal approximation, and with several charged legs and resonant propagators
in the final state the two are evaluated at kinematics that differ enough for
this ratio to become large. This is the variance cost of the squared-level
expansion discussed in \SecRef{sec:algorithm}, in its extreme form. The \ACRAIC
weight is the ratio of the coherent sum to the generator's own photon density
\EqRef{eq:crude}, both at the same photon kinematics, and it is positive at
$\mathcal O(\alpha)$. Its tails are far milder, and its statistical errors in
\TabRef{tab:4f} are $0.3$--$0.8\%$. An \oforder{\alpha^2} extension is possible
in principle, since \ACRAIC's tree-level $\bet{2}$ and real-virtual
construction do not depend on the final state, only the one-loop amplitude
does. For this charged-current topology we have experienced some numerical
instabilities, which is not unexpected for a \NNLOEW calculation for a
$2\rightarrow 4$ process, and we are investigating methods to alleviate them,
to be reported in a separate work.

\FloatBarrier
\section{Conclusions and outlook}\label{sec:conclusions}
We have presented \ACRAIC, an automated amplitude-level \YFS exponentiation for
general final states in \Sherpa. Of the ingredients for coherent exclusive
exponentiation only the coherent sum depends on the final state being a fermion
pair. We replace the initial/final assignment of photons by an assignment to
emission stages where the gauge invariance of every term requires each stage to
be charge neutral, and closure requires the stages to exhaust the charged legs
exactly once. The stages are derived from the radiating dipoles of the process,
with a resonance entering as a leg of both its production and its decay stage,
which for $W$ pairs would give the production and decay stages of the double-pole
picture. The remainders are tree-level helicity amplitudes with $N+2$ and $N+3$
legs, evaluated at the physical legs with partition-dependent propagator
shifts, which the \Comix recursion supplies directly. The weight is formed
against the generator's own photon density, the incoherent sum of the stage
eikonals on the post-emission legs, and is positive by construction at
$\mathcal O(\alpha)$. The
one-loop virtual enters through a per-helicity factor built from the loop
provider's amplitudes, whose Born-weighted average reproduces the squared-level
virtual exactly.

At photonic $\mathcal O(\alpha^2)$ \ACRAIC agrees with \kkmc and with
\YFSNLOEW to the levels of \SecRef{sec:validation}, with every difference
between the columns accounted for class by class. The
theory-nuisance-parameter bands of \SecRef{sec:uncertainty} cover the
missing photonic orders for the two example applications, the $Z$-pole line
shape (\SecRef{sec:nnlo}) and the Higgs recoil mass at FCC-ee
(\SecRef{sec:recoil}), apart from the backward fermion angle at the
$Z$ pole, where the $\mathcal O(\alpha)$ band is narrowly exceeded.

We expect \ACRAIC to be better suited to large-scale
production at FCC-ee, where the samples are far larger than any here and the
weights matter as much as the formal accuracy. Its weight is positive at
$\mathcal O(\alpha)$, with no negative weights against about $0.3\%$
for \YFSNLOEW for $\mu^+\mu^-$ at $240~\UGeV$ (\TabRef{tab:sign}), and for
four-lepton final states it avoids the single-event tails of the squared-level
weight (\SecRef{sec:4f}), so a large sample is not dominated by a few events.
Several extensions follow from the components described here. Radiation from
the $W$ decays in the generator would remove the residual radiative-return tail
of the electron charged-current channels. Once $W$ pairs are validated, the
difference between the one- and three-stage treatments will have to be quoted
as a scheme uncertainty.
A complete $\mathcal O(\alpha^2)$ matching further requires, in $\bet{0}^{(2)}$, the
two-loop $\gamma Z$ boxes and the two-loop weak corrections. The photonic part
of $\bet{0}^{(2)}$ is included, for $\ell^+\ell^-$ with the non-factorizable
boxes for photon exchange, and for $\gamma\gamma$ with the complete two-loop
hard function.

\begin{figure*}[t]
  \centering
  \begin{tikzpicture}[x=2.05cm, y=1.45cm, font=\small,
    cell/.style={draw=white, line width=1.2pt},
    lab/.style={align=center, inner sep=1pt}]
  \definecolor{llIn}{RGB}{88,190,110}
  \definecolor{llOpt}{RGB}{176,226,184}
  \definecolor{llPart}{RGB}{246,212,110}
  \definecolor{llNew}{RGB}{60,140,220}
  \definecolor{llYFS}{RGB}{232,242,252}
  \fill[llYFS] (0,0) rectangle (5.5,3);
  \newcommand{\llcell}[5]{
    \fill[#3] (#1,#2) rectangle ++(1,1);
    \draw[cell] (#1,#2) rectangle ++(1,1);
    \node[lab] at (#1+0.5,#2+0.66) {\normalsize #4};
    \node[lab, font=\scriptsize] at (#1+0.5,#2+0.27) {#5};}
  \newcommand{\llcellc}[4]{
    \fill[#3] (#1,#2) rectangle ++(1,1);
    \draw[cell] (#1,#2) rectangle ++(1,1);
    \node[lab] at (#1+0.5,#2+0.5) {\normalsize #4};}
  \llcellc{0}{0}{llIn}{$\bet{0}^{[0]}$}
  \llcellc{1}{0}{llIn}{$\bet{1}^{[0]}$}
  \llcellc{2}{0}{llIn}{$\bet{2}^{[0]}$}
  \llcell{3}{0}{llOpt}{$\bet{3}^{[0]}$}{option}
  \llcell{4}{0}{llOpt}{$\bet{4}^{[0]}$}{option}
  \llcellc{0}{1}{llIn}{$\bet{0}^{[1]}$}
  \llcellc{1}{1}{llIn}{$\bet{1}^{[1]}$}
  \llcell{2}{1}{llNew!55}{$\bet{2}^{[1]}$}{new}
  \fill[pattern=north east lines, pattern color=llNew!70] (3,1) rectangle ++(1,1);
  \draw[cell] (3,1) rectangle ++(1,1);
  \node[lab, fill=llYFS, inner sep=2pt] at (3.5,1.5) {\normalsize $\bet{3}^{[1]}$};
  \llcell{0}{2}{llPart}{$\bet{0}^{[2]}$}{photonic, EW}
  \foreach \n/\l in {1/2, 2/2, 3/2, 4/2, 4/1} {
    \draw[cell] (\n,\l) rectangle ++(1,1);
    \node[lab, gray] at (\n+0.5,\l+0.55) {\normalsize $\bet{\n}^{[\l]}$};}
  \foreach \l in {0,1,2} {
    \draw[cell] (5,\l) rectangle ++(0.5,1);
    \node[gray] at (5.25,\l+0.5) {$\cdots$};}
  \draw[very thick, dashed, black!70]
    (0,0) -- (0,2) -- (1,2) -- (1,1) -- (2,1) -- (2,0);
  \draw[very thick, black]
    (0,0) -- (0,3) -- (1,3) -- (1,2) -- (2,2) -- (2,1) -- (3,1) -- (3,0) -- cycle;
  \foreach \n in {0,...,4} {\node[below] at (\n+0.5,0) {$\n$};}
  \node[below] at (5.25,0) {$\cdots$};
  \foreach \l in {0,1,2} {\node[left] at (0,\l+0.5) {$\l$};}
  \node[below=12pt] at (2.75,0) {photons $n$ (legs)};
  \node[rotate=90] at (-0.28,1.5) {loops $\ell$};
  \draw[decorate, decoration={brace, amplitude=6pt}, very thick, llIn!60!black]
    (5.62,1.97) -- (5.62,0.03);
  \node[rotate=-90, font=\scriptsize\bfseries, llIn!50!black, align=center]
    at (5.86,1.0) {automated:\\any process};
  \draw[decorate, decoration={brace, amplitude=5pt}, very thick, llPart!60!black]
    (5.62,2.97) -- (5.62,2.03);
  \node[rotate=-90, font=\scriptsize, llPart!50!black, align=center]
    at (5.86,2.5) {process-\\specific};
  \begin{scope}[shift={(6.75,0)}]
    \newcommand{\lgcell}[4]{
      \fill[#3] (0.42*#1,0.75*#2) rectangle ++(0.42,0.75);
      \draw[cell] (0.42*#1,0.75*#2) rectangle ++(0.42,0.75);
      \node[lab, font=\scriptsize] at (0.42*#1+0.21,0.75*#2+0.375) {#4};}
    \fill[llYFS] (0,0) rectangle (1.26,3);
    \lgcell{0}{0}{llIn}{exact}
    \lgcell{1}{0}{llIn}{exact}
    \fill[white] (0.84,0) rectangle ++(0.42,0.75);
    \lgcell{0}{1}{llIn}{exact}
    \lgcell{1}{1}{llIn}{exact}
    \lgcell{2}{1}{llPart}{partial}
    \foreach \r in {2,3} {\foreach \c in {0,1,2} {\lgcell{\c}{\r}{llYFS}{soft}}}
    \draw[very thick] (0,0) -- (0,1.5) -- (1.26,1.5) -- (1.26,0.75) -- (0.84,0.75)
      -- (0.84,0) -- cycle;
    \node[left] at (0,0.375) {$\alpha$};
    \node[left] at (0,1.125) {$\alpha^2$};
    \node[left] at (0,1.875) {$\alpha^3$};
    \node[left] at (0,2.625) {$\alpha^{k\ge4}$};
    \node[below] at (0.21,0) {$L^k$};
    \node[below] at (0.63,0) {$L^{k-1}$};
    \node[below] at (1.05,0) {$L^{k-2}$};
    \node[below=12pt] at (0.63,0) {logarithms};
    \node[rotate=90] at (-0.42,1.5) {order $\alpha^k$};
  \end{scope}
  \begin{scope}[shift={(0,-1.15)}, font=\scriptsize]
    \newcommand{\llkey}[3]{\fill[#2] (#1,0) rectangle ++(0.13,0.16);
      \node[anchor=west] at (#1+0.15,0.08) {#3};}
    \llkey{0.00}{llIn}{in the $\mathcal O(\alpha^2)$ prediction, exact}
    \llkey{2.40}{llOpt}{exact, option}
    \llkey{3.45}{llNew!55}{exact, option (new)}
    \fill[pattern=north east lines, pattern color=llNew!70] (4.90,0) rectangle ++(0.13,0.16);
    \node[anchor=west] at (5.05,0.08) {feasible};
    \llkey{5.85}{llPart}{partial}
    \fill[llYFS] (0,-0.33) rectangle ++(0.13,0.16);
    \draw[black!30, line width=0.3pt] (0,-0.33) rectangle ++(0.13,0.16);
    \node[anchor=west] at (0.15,-0.25) {soft (infrared) logarithms only, from the YFS exponent, to all orders};
    \draw[very thick] (0,-0.58) -- ++(0.25,0);
    \node[anchor=west] at (0.27,-0.58) {$\mathcal O(\alpha^2)$ prediction};
    \draw[very thick, dashed, black!70] (1.75,-0.58) -- ++(0.25,0);
    \node[anchor=west] at (2.02,-0.58) {$\mathcal O(\alpha)$ prediction};
    \node[anchor=west] at (3.6,-0.58) {$L=\ln(s/m_e^2)$};
  \end{scope}
\end{tikzpicture}
  \caption{What \ACRAIC includes, by loops $\ell$ and photons $n$. Left: the
    residual $\bet{n}^{[\ell]}$ of \EqRef{eq:AJ}, contributing at
    $\mathcal O(\alpha^{n+\ell})$. The staircases enclose the
    $\mathcal O(\alpha)$ and $\mathcal O(\alpha^2)$ predictions of this paper.
    Right: the resulting logarithmic accuracy of the $\mathcal O(\alpha^2)$
    prediction in $\alpha^kL^m$.}
  \label{fig:loopslegs}
\end{figure*}
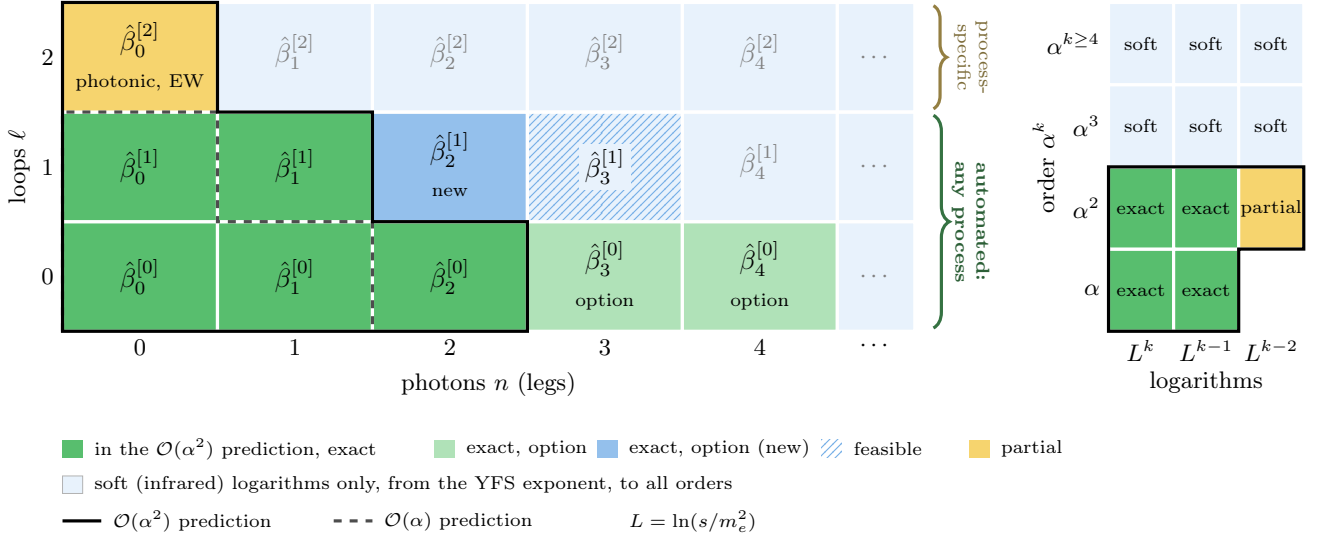

\FigRef{fig:loopslegs} summarizes which amplitudes enter, by loop order and
photon multiplicity, and the resulting logarithmic accuracy. The tree and
one-loop residuals are constructed automatically for any process from the
tree amplitudes of \Comix and the one-loop amplitudes of the loop provider.
$\bet{3}^{[0]}$, $\bet{4}^{[0]}$ and $\bet{2}^{[1]}$ are available as
run-time options beyond the $\mathcal O(\alpha^2)$ predictions validated
here, and $\bet{3}^{[1]}$ needs only the corresponding one-loop amplitude
once a provider returns it. The two-loop $\bet{0}^{[2]}$, the two-loop part
of $\bet{0}^{(2)}$, is the only process-specific ingredient, since it draws
on literature results rather than an automated amplitude.

 The initial-state photons of \ACRAIC are attached to the electron line, so the initial-state
  treatment can be combined in \Sherpa with its state-of-the-art description
  of QCD final states. This covers next-to-leading-order
  QCD matrix elements matched to parton showers~\cite{Hoeche:2011fd},
  multijet merging at next-to-leading order~\cite{Hoeche:2012yf}, including
  $e^+e^-\to$ hadrons~\cite{Gehrmann:2012yg}, and the \Sherpa parton
  showers~\cite{Schumann:2007mg,Hoche:2015sya,Herren:2022jej}. Photon
  emission from the quarks, its interplay with the QCD evolution and the
  initial-final interference in coloured processes then need a consistent
  treatment.

\begin{acknowledgments}
  A.P. thanks the other members of the \Sherpa collaboration for their work
  and dedication, which made this paper possible. A.P would also like to thank Ayres Freitas for many discussions
  on two-loop corrections and Jonas Lindert for continued support with \OpenLoops.
  The work of A.P.\ is supported by grant No.\
  2023/50/A/ST2/00224 of the National Science Centre (NCN), Poland.
  A.P gratefully acknowledges the Polish high-performance computing infrastructure
  PLGrid (HPC Center: ACK Cyfronet AGH) for providing computer facilities
  and support within computational grant no. PLG/2026/019132.
  A.P.\ thanks Claude (Anthropic) for assistance improving the
  presentation of this paper and for help testing the numerical implementation
  of the algorithm.
\end{acknowledgments}

\FloatBarrier
\appendix

\section{Conventions for the soft factors}\label{app:conventions}
The relative phases of the terms of \EqRef{eq:stage-partition-sum} are
physical, and an inconsistency between the conventions of the amplitudes and of
the soft factors is invisible to every squared quantity. We collect here the
conventions on which they depend.

\paragraph{Legs and charges.} All momenta are physical, with positive energy. The direction label $\theta_i$
of \EqRef{eq:theta} carries the crossing of the incoming legs, so that an
incoming particle of charge $Q_i$ enters the soft factors as an outgoing
particle of charge $-Q_i$. In terms of the sign $Z_i$ of
Ref.~\cite{Price:2025fiu}, $\theta_i=-Z_i$ and $Q_i\theta_i=-Z_iQ_i$ in every
soft factor. A resonance counted as a leg of two stages enters the production
stage with $\theta=+1$ and its decay stage with $\theta=-1$, at the same
momentum, so that its contributions cancel in \EqRef{eq:closure}.

\paragraph{Polarization vectors.} The polarization vectors $\varepsilon^{(\sigma)}(k)$ in
\EqRefs{eq:soft-factor}{eq:stage-soft-factor} are those of the external vector
currents of \Comix for an outgoing massless vector, constructed from its
spinors with the gauge reference vector $n=(1,0,1,0)$. The same vectors enter
$\Mn{1}$, so that $\sfac$, $\Mzero$ and $\Mn{1}$ share one convention.

\paragraph{Stage soft factors.} The stage factors \EqRef{eq:stage-soft-factor} are the restriction of the same
sum to the legs of the stage, with the same signs and normalization, so that
\EqRef{eq:closure} holds identically when the stages exhaust the charged legs.
The squared-level soft factor is \EqRef{eq:stilde-amplitude}.

\section{Beam energy spread}\label{app:bes}
We model the beam energy
spread (BES) by an independent Gaussian for each beam, the parametrization of
\kkmc~\cite{Jadach:2022mbe}. Beam $i$ has nominal
energy $E_0$ and relative spread $r$, so $\sigma_E=rE_0$, and we truncate the
Gaussian at $N$ standard deviations. For FCC-ee, $r$ ranges from
$0.13\%$ at the $Z$ pole to $0.165\%$ at the $ZH$ energy, including
beamstrahlung~\cite{FCC:2018evy}. The beam momenta are built at the maximum
energy $E_{\max}=E_0(1+Nr)$, and each event samples the energy fractions
$x_i=E_i/E_{\max}$ with the density
\begin{align}
  f(x_i) = \frac{1}{\sqrt{2\pi}\,\sigma_x\,\mathrm{erf}(N/\sqrt2)}\nonumber\,
           \exp\!\left[-\frac{(x_i-x_0)^2}{2\sigma_x^2}\right],\\
  \qquad x_0=\frac{1}{1+Nr},\quad \sigma_x = r\,x_0 ,
  \label{eq:bes}
\end{align}
on $|x_i-x_0|\le N\sigma_x$, i.e.\ centred on $E_0$ with width $\sigma_E$ and
normalized to unity on the truncated interval. For uncorrelated beams the
density of the pair is the product $f(x_1)f(x_2)$. A correlation $\rho$ of the
two beam energies is included by the ratio of the correlated two-dimensional
Gaussian to that product,
\begin{equation}
  \begin{split}
  f(x_1,x_2) ={}& \frac{\mathcal N_\rho}{\sqrt{1-\rho^2}}\, f(x_1)f(x_2)\\
    &\times\exp\!\left[-\frac{\rho^2(d_1^2+d_2^2)-2\rho\,d_1d_2}
    {2(1-\rho^2)}\right],
  \end{split}
  \label{eq:bes-rho}
\end{equation}
with $d_i=(x_i-x_0)/\sigma_x$, where $\mathcal N_\rho$ renormalizes the correlated density to unity on the
truncated square $|d_i|\le N$. We take the default $N=4$ and the probability
outside the truncation is $6\cdot10^{-5}$.

An event therefore has its own centre-of-mass energy
$s=(p_1+p_2)^2$ which no longer means $E_1 = E_2$. Every
quantity of both columns that depends on the energy is evaluated with the
$s$ of the event, not with the nominal $4E_0^2$:
\begin{itemize}
  \item The initial-state dipole is built on the beams of the event, and the
    initial-state photons, their soft cutoff $\varepsilon\sqrt s/2$ and the
    reduced energy $s'$ are generated with respect to them.
  \item The form factors of all dipoles, the soft scale of the initial-final
    exponent, the final-state cutoff, the virtual subtraction and the soft cuts
    of the matching are evaluated at the event's $s$.
  \item \ACRAIC evaluates the partition sum and the crude density in the
    centre-of-mass frame of the event.
\end{itemize}
The event is generated in that frame and boosted to the laboratory along the
beam axis, together with all photons. Only the analysis refers to the nominal
energy.

\bibliographystyle{apsrev4-2}
\bibliography{main}

\end{document}